# Observability Engineering: From Measurement to Information Generation in Active Sensing Systems

Pin-Han Ho and Limei Peng

***Abstract*—Sensing systems across radar, robotics, wireless communications, and scientific instrumentation are undergoing a fundamental transition from passive measurement toward interaction-driven perception, where performance depends not only on measurement quality, but also on how sensing actions are selected and coordinated over time. In such systems, trajectories, viewpoints, waveforms, frequencies, spatial configurations, and interaction policies actively shape which aspects of the environment become distinguishable.**

**This article develops a four-stage observability-engineering framework for interaction-driven sensing. The framework integrates physical modeling, information geometry, symmetry reduction, cross-system correspondence, and observable world models into a unified engineering perspective. It progresses from physics-constrained observation families and trajectory-induced Fisher information geometry, to symmetry-reduced quotient-space observability, cross-system information equivalence, and observable world-model construction. Within this perspective, sensing actions induce information structures that govern local distinguishability, while physical symmetries define equivalence classes of states that remain fundamentally indistinguishable.**

**The proposed framework explains several common phenomena in modern sensing systems. Physically different architectures can exhibit comparable sensing capability when they induce similar quotient-space information structures, whereas insufficient action diversity can lead to blind directions, ill-conditioned estimation, and false identifiability. The framework also provides a quotient-space language for comparing heterogeneous sensing architectures beyond hardware-level descriptors, while clarifying how different forms of sensing diversity contribute to distinguishability.**

**The framework is developed primarily through high-frequency electromagnetic sensing, where geometric-optics approximations provide an analytically interpretable physical realization. The broader observability-engineering perspective, however, does not depend on geometric optics itself; it requires an action-indexed observation family and a task-relevant distinguishability measure. From this viewpoint, observability is not a passive property of sensor measurements alone, but an engineered outcome of action–physics coupling under physical and symmetry constraints.**

## I. Introduction

Sensing has historically been formulated as a passive measurement process, in which a fixed observation operator maps unknown environment parameters to measured data. This viewpoint underlies classical radar, imaging, and statistical estimation theory [1, 2, 3, 4, 5, 6, 7, 8, 9, 10, 11], where the sensing interface is assumed to be given and the central problem is inversion under noise and model uncertainty. Fundamental limits, including resolution, identifiability, and estimation accuracy, are therefore derived under fixed sensing configurations and are often characterized by bandwidth, aperture size, or signal-to-noise ratio [12, 13, 14].

Pin-Han Ho is with University of Waterloo, Canada, and Limei Peng is with Kyungpook National University, Korea. Both are also with Shenzhen Institute for Advanced Study, UESTC.

**This article offers a different perspective:**

*Observability is not determined by sensors alone, but emerges from sensing trajectories executed within a physically constrained control space.*

More precisely, controllable sensing trajectories induce a statistical distinguishability structure over observable states. The article later formalizes this structure using Fisher information and symmetry-reduced quotient representations. Sensing systems that induce equivalent information geometry on the observable quotient space can therefore exhibit equivalent local distinguishability even when their hardware implementations differ.

Modern sensing systems increasingly operate in regimes where the observation process itself is controllable. Advances in synthetic aperture radar (SAR), computational imaging, reconfigurable electromagnetic environments, and integrated sensing and communication (ISAC) enable platforms to adapt frequency, viewpoint, waveform, polarization, and even the propagation environment [15, 16, 17, 18, 19, 20, 21, 22, 23, 24, 25, 26, 27]. In parallel, robotics and active perception have long emphasized that sensing performance depends critically on how measurements are selected and coordinated over time [28, 29, 30, 31, 32, 33]. These developments point toward a shift from *measurement as acquisition* to *measurement as interaction*.

When sensing actions are controllable, physically distinct architectures may generate equivalent information when operated appropriately, while identical hardware may yield different information under different strategies [34, 35].

**A central question emerges:** *How do controllable sensing actions generate information about the physical world, and how can that information be represented, compared, and designed?*

*Why this matters for engineering practice.* Traditional approaches treat sensing as a hardware problem: selecting a radar, an antenna array, or a camera, and then deriving performance bounds. In modern systems such as cognitive radar, autonomous platforms, and ISAC, however, the primary design freedom increasingly lies in how the hardware is operated.

This perspective provides a unified lens for questions such as *whether* limited resources should be allocated to wider bandwidth or greater spatial diversity, *why* estimators may fail despite advanced hardware configurations, and *whether* a MIMO radar can reproduce the observability of a moving monostatic platform using fewer channels. The answer lies in the *information geometry* induced by sensing trajectories, namely a structure shaped through sensing design, diagnosed through trajectory-level indicators, and compared across architectures.

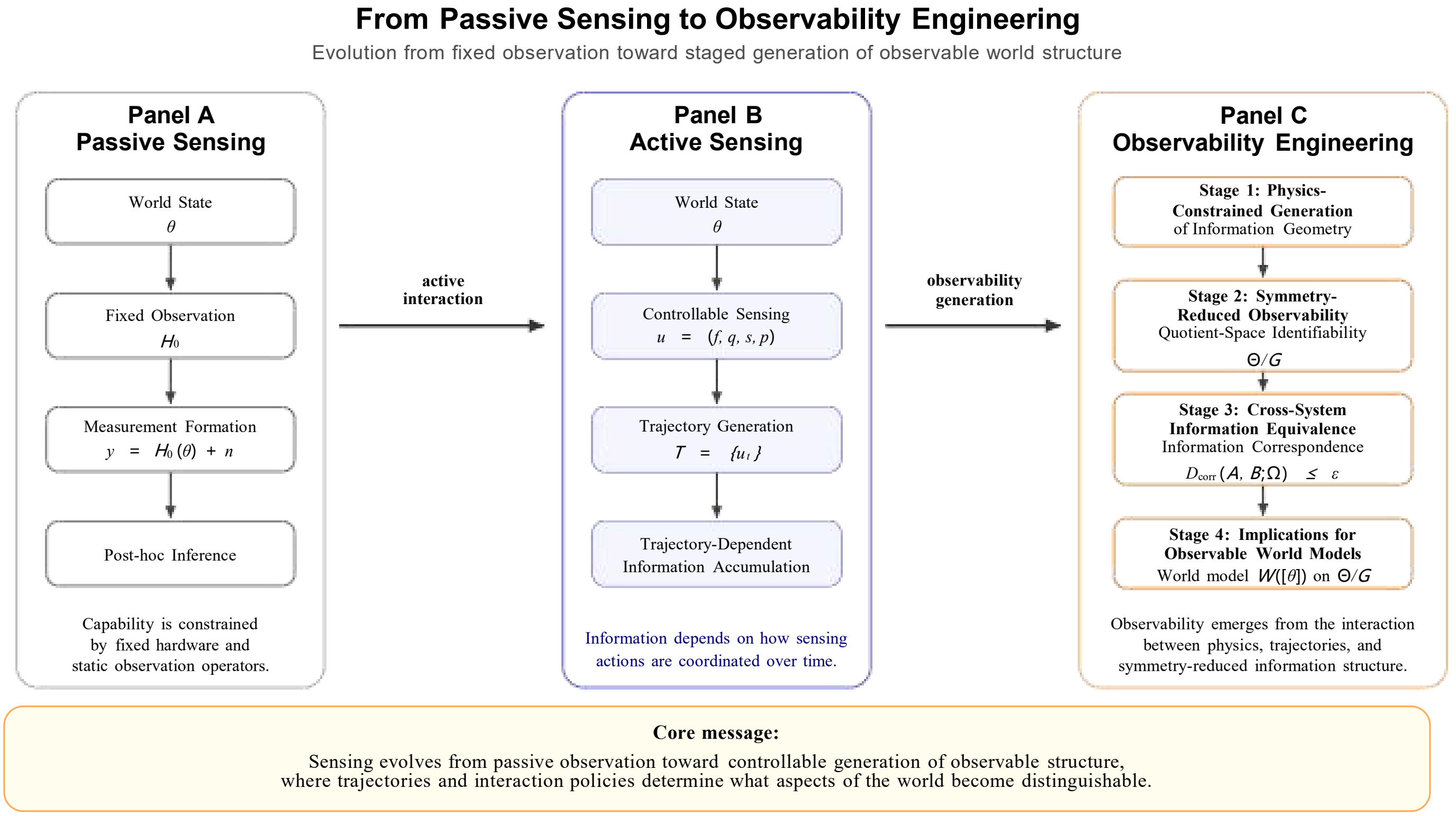

Fig. 1. **From passive sensing to observability engineering.** Panel A represents classical passive sensing based on fixed observation operators and post-hoc inference. Panel B introduces controllable sensing actions and trajectory-dependent information acquisition. Panel C further elevates sensing into observability engineering, where controllable interaction generates information geometry, symmetry-reduced observability, cross-system information equivalence, and observable world models. Cross-panel arrows indicate conceptual progression rather than direct signal flow: passive sensing motivates controllable interaction, and controllable interaction is formalized as observability engineering.

Significant advances have been made across estimation theory, control-theoretic observability, inverse problems, experimental design, and information geometry [6, 7, 36, 37, 38, 39, 40, 41, 42, 43]. Yet they share a common limitation: the role of controllable sensing actions in shaping distinguishability remains implicit. This gap motivates a shift from analyzing inference under given measurements to understanding sensing as a process of *information generation*.

This article introduces an engineering perspective termed *observability engineering*, which interprets sensing as the deliberate generation of information through controllable observation processes governed by Maxwell's equations [44, 45, 46, 47, 48, 49, 50]. Rather than treating sensing systems as fixed measurement devices, the perspective views them as *generative mechanisms* that induce information geometry through trajectories in a controllable action space.

Electromagnetic sensing is used in this article as the canonical physical realization because it exposes, in an analytically interpretable form, the interaction among controllable actions, wave propagation, and statistical distinguishability. The framework is therefore grounded in electromagnetism, particularly in embodied sensing systems where actions are physically realized and directly coupled to the observation process. However, the broader observability-engineering perspective does not depend on geometric optics itself; what it requires is an action-indexed observation family together with a task-relevant measure of distinguishability. Geometric-optics approximations are used only when they provide useful structure for interpreting the induced information geometry.

**Terminology Clarification: What "Observability" Means Here**

In classical control theory, observability refers to state reconstructability from measurement histories. In this article, *observability engineering* is used in an engineering and trajectory-level sense: the design of sensing actions so that the induced information structure sufficiently excites observable directions on $\Theta/G$. The framework complements rather than replaces control-theoretic observability.

This article does not aim to replace classical observability, estimation theory, or experimental design. Rather, it provides a structural engineering viewpoint that organizes these tools around a common question: how sensing actions generate distinguishability under physical and symmetry constraints.

From a unified perspective, the four stages in Table I are not separate modules, but different aspects of a single information-generation process. Controllable actions shape distinguisha-

TABLE I

OBSERVABILITY ENGINEERING: A STAGED FRAMEWORK ALIGNED WITH FIG. 1. THE PROPOSED FRAMEWORK IS ORGANIZED AS A PROGRESSION FROM PHYSICAL INTERACTION TO OBSERVABLE WORLD MODELS . THE TABLE SUMMARIZES THE CANONICAL CONCEPTS , INTERPRETATIONS , AND DESIGN IMPLICATIONS ASSOCIATED WITH EACH STAGE .

| Stage | Canonical Concept | Interpretation | Implication for Design |
|---|---|---|---|
| **Stage 1: Physics-Constrained Generation of Information Geometry** | | | |
| 1-1 Maxwell Observation Family | Parameterized observation operators | Physical laws define a controllable observation family $y = H(u;\theta)+n$. | Different architectures can be viewed as realizations of a shared physics-constrained observation family. |
| 1-2 Trajectory-Induced Information Generation | Action-indexed sensing trajectories | A sensing trajectory $\mathcal{T} = \{u_t\}_{t=1}^{T}$ determines how observations are generated and accumulated. | Information quality depends on action sequencing, not static hardware alone. |
| 1-3 Fisher Information Geometry | Local distinguishability structure | Trajectory-induced sensitivities define Fisher information $\mathbf{I}(\mathcal{T};\theta)$ in the ambient parameter space. | Design should excite informative directions and avoid ill-conditioned information structures. |
| **Stage 2: Symmetry-Reduced Observability** | | | |
| 2-1 Quotient Representation of Observable States | Observable equivalence classes | Physical symmetries define indistinguishable states $[\theta] \in \Theta/G$. | Inference should be formulated on observable equivalence classes rather than redundant ambient coordinates. |
| 2-2 Projected Fisher Information on the Quotient | Quotient-space distinguishability | Ambient Fisher information is projected onto directions transverse to symmetry orbits, yielding $\mathbf{I}_Q$. | Design should evaluate excitation only on quotient-identifiable directions. |
| **Stage 3: Cross-System Information Equivalence** | | | |
| 3 Cross-System Information Equivalence | Cross-platform information structure | Heterogeneous sensing systems are compared through their induced quotient-space information structures. | Systems should be compared by information equivalence, not hardware similarity. |
| **Stage 4: Implications for Observable World Models** | | | |
| 4 Observable World Models | Task-relevant observable representation | World models are constructed on observable states $[\theta] \in \Theta/G$, with uncertainty structured by $\mathbf{I}_Q^{-1}$. | Encode what is distinguishable and use quotient-space uncertainty to guide future sensing actions. |

bility, symmetries define observable equivalence classes, and world models constructed on the quotient space should remain consistent with the resulting distinguishability structure. The remainder of the article develops these stages in sequence.

The resulting perspective leads to several concrete conceptual and engineering contributions.

*a) Contributions:* Specifically, this article contributes:

1) A physics-grounded interpretation of sensing as a family of controllable observation processes;
2) A trajectory-based perspective on information generation through Fisher-induced geometry;
3) A symmetry-aware formulation of identifiability via observable equivalence classes on $\Theta/\mathcal{G}$;
4) A system-level lens for comparing heterogeneous sensing architectures through the information structures they induce on the observable quotient space.

*b) Scope and limitations:* The perspective is most directly applicable in high-frequency regimes where wave propagation admits geometric-optics approximations and scattering remains weak to moderate [15, 51, 52]. Strong multiple scattering, low-frequency regimes, and fully nonlinear materials are not treated in full generality. Fisher information provides a local measure of distinguishability but does not capture non-Gaussian or multi-modal inference behavior.

Fig. 1 summarizes the proposed paradigm shift. Panel A illustrates classical passive sensing. Panel B introduces controllable sensing actions and trajectory design. Panel C outlines the staged progression from Maxwell observation families through quotient-space identifiability to observable world model. Table I and Table II organize the conceptual structure.

## II. Stage 1: Physics-Constrained Generation of Information Geometry

This section develops Stage 1 of the proposed framework, namely **physics-constrained generation of information geometry** using electromagnetic sensing as a canonical realization. The goal is not to derive Maxwell solutions in full generality, but to illustrate how controllable sensing trajectories generate distinguishability through physics-constrained observation families.

Within the electromagnetic setting, Maxwell-based models offer a physically interpretable structure where phase-dominated geometric sensitivity and amplitude-based material response naturally emerge. These patterns illustrate general principles governing how sensing actions shape distinguishability.

### A. Stage 1-1: Maxwell Observation Family

This subsection develops the first internal stage of Stage 1: the Maxwell observation family underlying physics-constrained sensing interactions. Within the proposed perspective, sensing is interpreted not as a fixed measurement process, but as a controllable family of physical interactions governed by admissible electromagnetic observation operators.

TABLE II
POSITIONING OF THE PROPOSED FRAMEWORK RELATIVE TO EXISTING PARADIGMS

| Paradigm | Core Question | Limitation | This Framework's Addition |
|---|---|---|---|
| Control-Theoretic Observability [37] | Is the state reconstructible? | Global, noise-free | Trajectory-level, noise-aware |
| Fisher Information / CRB [7] | How accurate can estimation be? | Local, fixed model | Trajectory-induced geometry on quotient |
| Optimal Experimental Design [40] | Which measurements to take? | Predefined model class | Shaping quotient-space information |
| Active Perception / SLAM [28, 31] | Where to move to improve estimates? | Heuristic diversity | Unified information-geometric diagnostic |

We consider a parameterized observation family

$$y = H(u;\theta) + n, \tag{1}$$

where $\theta \in \Theta$ denotes the environment state, $u \in \mathcal{U}$ represents controllable sensing actions, $H(u;\theta)$ is the corresponding observation operator, and $n$ denotes measurement noise. The control variable $u$ may include sensing frequency, spatial configuration, waveform parameters, polarization states, platform motion, or other controllable sensing dimensions [53, 54, 55].

Under this formulation, sensing systems are interpreted as mechanisms for generating observations through controllable physical interaction. Different sensing architectures may often be interpreted as different realizations of a shared physics-constrained observation family, despite substantial implementation differences [34, 56, 57].

The control variable $u$ will later be organized into sensing trajectories $\mathcal{T} = \{u_t\}_{t=1}^{T}$, as developed in Stage 1-2. This trajectory-oriented interpretation is consistent with active perception and adaptive sensing perspectives, where measurement selection and action design shape the information obtained from the environment [28, 58, 59]. At this stage, the key point is that physical laws define not a single fixed measurement operator, but an admissible family of observation operators indexed by controllable sensing actions.

Electromagnetic sensing is governed by Maxwell's equations. For linear isotropic media in the frequency domain,

$$\nabla \times \mathbf{E} = -j\omega\mu\mathbf{H}, \qquad \nabla \times \mathbf{H} = (\sigma + j\omega\epsilon)\mathbf{E} + \mathbf{J}_{\mathrm{src}}, \tag{2}$$

where $\epsilon$, $\mu$, and $\sigma$ denote the permittivity, permeability, and conductivity, respectively [44, 45, 46]. Within the present framework, these equations define the admissible structure of physical interaction rather than objects to be solved analytically in full generality [3, 60].

To expose design-relevant structure, we adopt a high-frequency asymptotic representation. Under standard geometric-optics conditions ($kL \gg 1$),

$$\mathbf{E}(\mathbf{r};k) \approx \sum_{p\in\mathcal{P}(u)} A_p(\theta,u)\exp\big(-jk\,S_p(\theta_{\mathrm{g}},u)\big), \tag{3}$$

where $\mathcal{P}(u)$ denotes the set of admissible propagation paths, $A_p$ represents amplitude-dependent effects, and $S_p$ denotes geometric optical path length [49, 61, 45, 62].

This representation reveals an important structural asymmetry. Geometric perturbations predominantly affect phase terms through the factor $kS_p$, producing sensitivity that scales approximately with wavelength-normalized path variation. Material perturbations, by contrast, primarily influence amplitude terms and often remain comparatively weaker in high-frequency regimes [63, 51]. As a consequence, different parameter subspaces generate distinguishability with substantially different scaling behavior [64, 63, 51].

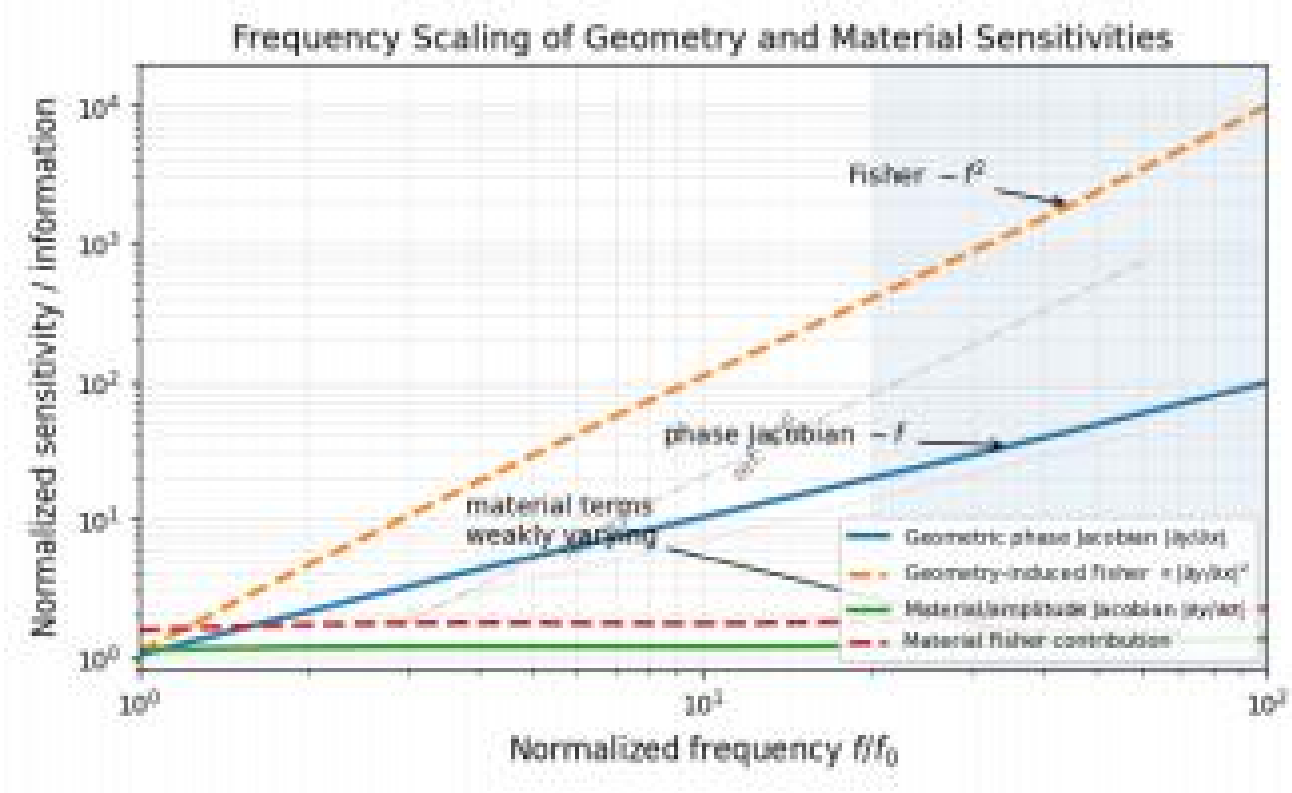

Fig. 2. High-frequency scaling of geometric and material sensitivities. Geometric perturbations primarily affect phase terms and therefore produce Jacobian growth approximately proportional to frequency, leading to Fisher-information growth approximately proportional to $f^2$. Material or amplitude-related sensitivity varies more weakly across the same frequency range. The shaded region indicates the high-frequency asymptotic regime where phase-dominated geometric sensitivity becomes increasingly prominent.

This phase–amplitude asymmetry provides an intuitive explanation for anisotropic information generation across parameter dimensions. As illustrated in Fig. 2, geometric phase sensitivity exhibits approximately linear frequency scaling, while the corresponding Fisher information grows approximately quadratically with frequency. Material sensitivity, by contrast, varies more weakly across the same range. The figure is intended to illustrate the asymptotic scaling trend rather than represent a specific calibrated electromagnetic scenario.

A partial suppression of cross-coupling terms may occur in high-frequency regimes, where rapidly oscillating phase terms can lead to cancellation effects consistent with stationary-phase arguments. This motivates, but does not universally guarantee, an approximately decoupled sensitivity structure under weak-to-moderate multiple-scattering conditions [49, 45, 65].

The phase–amplitude separation discussed above should be interpreted as a high-frequency structural heuristic rather than a universal property. Its validity is governed primarily by the electrical size

$$kL = \frac{2\pi L}{\lambda},$$

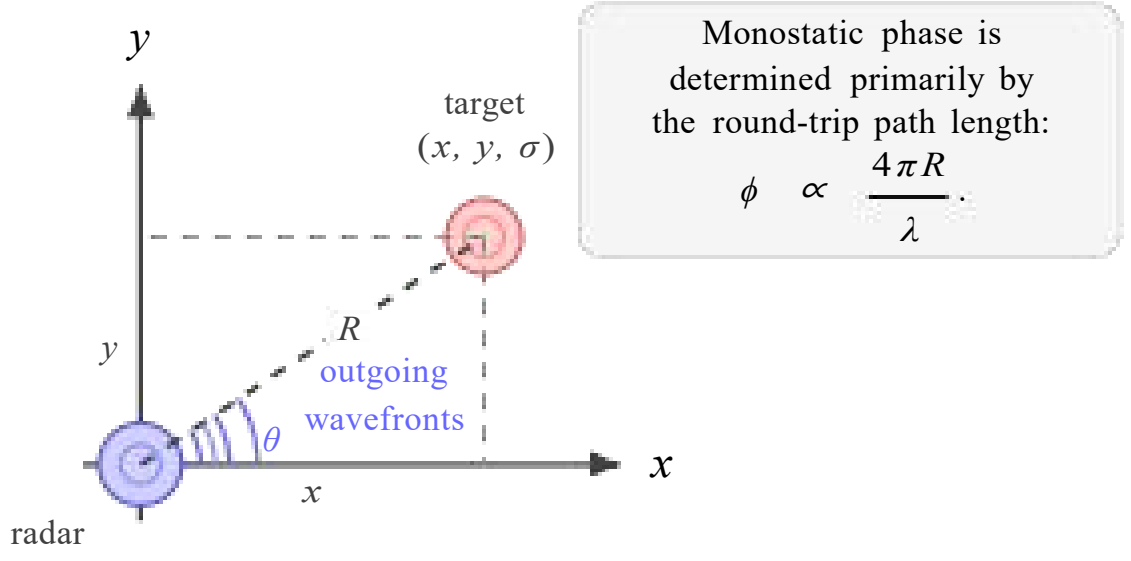

Fig. 3. Monostatic radar observing a point target. The round-trip range $R$ determines the received phase, while the target location can be represented by its Cartesian coordinates $(x, y)$ and reflectivity $\sigma$.

where $L$ denotes a characteristic scene dimension. The phase-dominant interpretation becomes increasingly accurate when $kL \gg 1$ and geometric-optics assumptions hold, while multiple scattering remains weak to moderate [49, 12, 45, 61]. In low-frequency or resonant regimes where $kL \approx 1$, wavelength-scale interactions become significant, geometric phase dominance may disappear, and amplitude–phase coupling can become strong [64, 66, 67, 62].

The simplified structure may also break down in dense multiple-scattering environments, strongly resonant structures, nonlinear media, or settings where higher-order scattering contributes non-negligibly to the received field energy [65, 68, 69, 45]. In such cases, the full forward model should be used to compute the information structure directly, rather than relying on geometric-optics intuition or phase–amplitude separation [63, 51, 70].

These limitations do not invalidate the observability-engineering framework itself. Rather, they indicate when simplified asymptotic interpretations should be replaced by numerically evaluated information models. The quotient-space formulation remains useful for identifying symmetry-induced equivalence [71, 72], but the Fisher-based local information geometry should be complemented, or in some cases replaced, by global discrepancy measures such as Kullback–Leibler divergence, Wasserstein distance, or task-specific Bayesian risk when nonlinearities, nonlocal ambiguities, or strongly multimodal posteriors dominate [73, 74, 42, 75].

Fig. 3 illustrates a monostatic radar observing a point target, where the received phase is governed primarily by round-trip propagation distance. The information geometry generated in Stage 1 will later be interpreted under symmetry reduction in Section III.

### B. Stage 1-2: Trajectory-Induced Information Generation

The control space $\mathcal{U}$ includes configuration variables (spatial arrangement, platform pose) and action variables (frequency, waveform, polarization) [53, 26, 76, 54]. A trajectory $\mathcal{T} = \{u_t\}_{t=1}^{T}$ corresponds to a path in $\mathcal{U}$, inducing a sequence of observation operators (1). In embodied systems, the control space is physically constrained, shaping the observation family.

Architectural variability—MIMO arrays, frequency-diverse apertures, reconfigurable surfaces, embodied platforms—reflects differences in the reachable control space, not fundamentally different governing equations [34, 56, 19, 20]:

- **MIMO/phased arrays**: spatial diversity via multiple channels and beamforming [77, 78];
- **Frequency-diverse systems**: spectral diversity via waveform design [57, 79];
- **Reconfigurable environments**: programmable propagation via metasurfaces [21, 22];
- **Embodied platforms**: diversity through motion and viewpoint variation [15, 17].

These are different realizations of the same observation family, providing a common abstraction for sensing diversity.

> *Sensing design is not inversion of a fixed operator, but generation of distinguishability through controllable observation families Eq.* (1).

Fig. 4 summarizes Stage 1 as the physics-constrained generation of information geometry.

### C. Stage 1-3: Fisher Information Geometry

The environment is represented by $\theta \in \Theta$, distinguishing geometric parameters $\theta_{\mathrm{geom}}$ (position, shape) from material parameters $\theta_{\mathrm{mat}}$ (reflectivity, permittivity) [64, 63, 70].

Local sensitivity is described by the Jacobian $\mathbf{J}(u;\theta) = \partial H(u;\theta)/\partial\theta$. Under circularly symmetric complex Gaussian noise $n \sim \mathcal{CN}(0, \sigma^2\mathbf{I})$, the Fisher information for a single action is

$$\mathbf{I}(u;\theta) = \frac{2}{\sigma^2}\Re\{\mathbf{J}^H\mathbf{J}\}, \tag{4}$$

consistent with classical estimation theory [7, 80, 11, 10]. For a trajectory $\mathcal{T}$, information accumulates as

$$\mathbf{I}(\mathcal{T};\theta) = \sum_{t=1}^{T}\mathbf{I}(u_t;\theta), \tag{5}$$

under independence assumptions [7, 36]. In correlated or adaptive settings, this becomes a non-additive functional [81].

*Fisher information as a local surrogate:* Throughout this article, Fisher information is used as a representation of local distinguishability induced by sensing actions. This choice is motivated by its analytical tractability and its direct connection to estimation accuracy under standard regularity conditions [82, 83].

However, it is important to emphasize that Fisher information provides a *local, second-order approximation* of statistical distinguishability [43]. In general, distinguishability is more fundamentally characterized by divergence measures such as the Kullback–Leibler divergence, particularly in non-Gaussian or multi-modal settings [75, 73].

The role of Fisher information in this work is therefore interpretative rather than exhaustive: it serves as a tractable proxy that reveals how sensing trajectories shape local information geometry, while acknowledging that global inference behavior may exhibit richer structure beyond this approximation. This perspective aligns with common engineering practice, where Fisher information is widely used as a design proxy despite its known limitations [40, 41].

TABLE III
PHASE VS. AMPLITUDE SENSITIVITY: A DESIGN HEURISTIC FOR HOW PARAMETERS AFFECT MEASUREMENTS

| Parameter Type | Measurement Domain | Observed Scaling with Frequency |
|---|---|---|
| Geometric $(x, y)$ | Phase | Often $\alpha f$ (Jacobian) / $\alpha f_2$ (Fisher) |
| Material $\sigma$ | Amplitude | Typically $O(1)$ within operational bands |

Throughout the paper, we denote Fisher information matrices by $\mathbf{I}$. Fisher information is interpreted here in an engineering sense: a representation of how controllable actions generate distinguishability. The quadratic form $ds^2 = d\theta^T \mathbf{I}(\theta) d\theta$ provides a local measure of distinguishability for subsequent analysis [84, 85].

*Physical structure of the induced Fisher geometry:* The high-frequency representation in Eq. (3) also explains why the Fisher geometry induced by electromagnetic sensing trajectories is often anisotropic [63, 51, 64]. Geometric perturbations primarily affect phase and therefore scale with wavenumber, whereas material perturbations often enter through weaker amplitude or scattering variations. Thus, trajectory-induced Fisher information may be strongly unbalanced across geometric and material subspaces. This should be viewed as a high-frequency design heuristic, not a universal theorem; outside geometric-optics or weak-scattering regimes, the full forward model should be used to compute the information structure.

## III. STAGE 2: SYMMETRY-REDUCED OBSERVABILITY

This section develops Stage 2 of the framework: *symmetry-reduced observability* (Fig. 1). The canonical concept associated with this stage is quotient-space identifiability. While Stage 1 generates a physics-constrained information geometry through Maxwell observation families, sensing trajectories, and Fisher information, Stage 2 identifies which directions correspond to observable quantities after symmetry reduction and separates intrinsic symmetry-induced degeneracy from design-dependent excitation effects.

The main role of Stage 2 is not to introduce another Fisher criterion, but to define the state space on which distinguishability should be evaluated. Fisher information enters only after symmetry-induced equivalence has determined the observable quotient structure.

### A. From Observability to Quotient Identifiability

Before introducing quotient-space identifiability, it is useful to clarify how the term *observability* is used in this article. The term carries distinct meanings across control theory, statistical estimation, and sensing-system design [37, 38]. Stage 2 connects these meanings by separating local distinguishability from intrinsic symmetry-induced non-identifiability.

Three related but distinct notions are relevant in the present framework.

*(i) Control-theoretic observability:* This refers to the ability to reconstruct system state from measurements over time, usually formulated as a global property in a noise-free dynamical-system setting [37, 86, 38, 87]. Classical results establish that observability depends on the pair $(C, A)$ in linear systems, with extensions to nonlinear systems through Lie-algebraic conditions [39, 88, 89].

*(ii) Fisher-based distinguishability:* This refers to a local measure of parameter sensitivity and estimation accuracy, depending on the true parameter value and sensing configuration [7, 6, 36]. The Fisher information matrix provides a local Riemannian metric on the parameter manifold [84, 43].

*(iii) Interaction-level observability:* In the engineering sense adopted in this article, observability refers to the ability of controllable sensing interactions to generate informative and distinguishable observations under finite samples and noise. This notion is captured operationally by Fisher-induced information structure evaluated on the observable quotient space [90, 91].

These notions are related but not equivalent. Control-theoretic observability addresses state reconstructability under dynamical evolution; Fisher information characterizes local statistical distinguishability; and interaction-level observability concerns whether controllable sensing actions sufficiently excite the directions that remain meaningful after symmetry reduction.

The Fisher information geometry generated in Stage 1 characterizes local distinguishability in the ambient parameter space. However, ambient Fisher information alone cannot distinguish between two qualitatively different sources of degeneracy: directions that are intrinsically unobservable because of physical symmetries, and directions that remain weakly excited because of insufficient sensing diversity [71, 92].

This distinction motivates the transition from ambient Fisher geometry to quotient-space identifiability. Symmetry-induced degeneracy should be removed through quotienting, while excitation-induced degeneracy should remain visible as poor conditioning or rank deficiency on the quotient-identifiable subspace [72, 93]. The quotient space $\Theta/\mathcal{G}$ therefore provides a geometric representation in which intrinsic indistinguishability can be separated from design-dependent excitation effects.

A key observation is that, in the ambient parameter space, both symmetry-induced indistinguishability and excitation-induced degeneracy appear as directions of low or vanishing sensitivity in the Fisher information matrix. From the ambient Fisher geometry alone, these two effects are generally inseparable because both manifest through reduced local curvature of the likelihood function [71]. As a result, analyses performed directly in the ambient parameter space may conflate intrinsic non-identifiability with insufficient excitation [92].

This distinction has important implications for observability analysis. Any observability analysis that does not explicitly account for symmetry-induced equivalence cannot, in general, distinguish intrinsic non-identifiability from insufficient excitation and is therefore structurally incomplete [94]. Quotient-space formulations provide the geometric structure necessary to separate these effects and define observability on the quotient-identifiable subspace.

TABLE IV

KEY QUANTITIES DEFINING THE OBSERVABILITY-ENGINEERING FRAMEWORK. THE TABLE IS ORGANIZED ACCORDING TO THE FOUR-STAGE STRUCTURE IN FIG. 1 AND TABLE I, WITH OPERATIONAL DIAGNOSTICS LISTED SEPARATELY.

| Symbol | Role in the Framework | First Appearance |
|---|---|---|
| **Stage 1: Physics-Constrained Generation of Information Geometry** | | |
| $u$ | Controllable sensing action, such as frequency, viewpoint, waveform, polarization, platform pose, or reconfigurable state. | Section II |
| $\mathcal{U}$ | Control space of admissible sensing actions constrained by the physical sensing system and operating environment. | Section II |
| $\mathcal{H}(u;\theta)$ | Physics-constrained observation operator indexed by sensing action $u$, defining the observation family $y = \mathcal{H}(u;\theta) + n$. | Eq. (1) |
| $\mathcal{T}$ | Sensing trajectory $\{u_t\}_{t=1}^{T}$, representing a sequence of controllable sensing actions that generates information through interaction with the environment. | Section II |
| $\mathbf{J}(u;\theta)$ | Local sensitivity matrix of the observation operator with respect to the environment state. | Eq. (4) |
| $\mathbf{I}(u;\theta)$ | Fisher information induced by a single sensing action, representing local distinguishability generated by one action-indexed observation. | Eq. (4) |
| $\mathbf{I}(\mathcal{T};\theta)$ | Fisher information accumulated along $\mathcal{T}$, defining the local distinguishability structure induced by sensing actions in the ambient parameter space. | Eq. (5) |
| **Stage 2: Symmetry-Reduced Observability** | | |
| $\mathcal{G}$ | Symmetry group acting on $\Theta$, defining transformations that leave the observation distribution invariant. | Eq. (6) |
| $[\theta]$ | Equivalence class of $\theta$ under the action of $\mathcal{G}$, representing a physically indistinguishable observable state. | Section III |
| $\Theta/\mathcal{G}$ | Observable quotient space where identifiability is defined after removing symmetry-induced redundancy. | Section III |
| $\mathcal{O}_\theta$ | Symmetry orbit of $\theta$, collecting all states related to $\theta$ by transformations in $\mathcal{G}$. | Section III |
| $\mathbf{P}_\perp(\theta)$ | Projection onto directions transverse to the symmetry orbit, used to isolate quotient-identifiable directions. | Eq. (8) |
| $\mathbf{I}_Q(\theta)$ | Projected Fisher information on the quotient space, obtained by removing directions tangent to the symmetry orbit. | Eq. (8) |
| $\mathbf{I}_Q(\mathcal{T};\theta)$ | Trajectory-accumulated projected Fisher information, measuring how a sensing trajectory excites quotient-identifiable directions. | Eq. (9) |
| **Stage 3: Cross-System Information Equivalence** | | |
| $\mathcal{A}, \mathcal{B}$ | Sensing architectures or systems being compared through their induced quotient-space information structures. | Section IV |
| $\Omega$ | Task-relevant region of the observable quotient space over which sensing systems are compared. | Section IV |
| $\mu$ | Measure over $\Omega$, such as a uniform, prior-based, or task-weighted measure used in regional information correspondence. | Eq. (11) |
| $D_{\mathrm{corr}}$ | Regional information-correspondence discrepancy between sensing architectures $\mathcal{A}$ and $\mathcal{B}$, based on the quotient-space information structures they induce over $\Omega$. | Eq. (11) |
| $\tilde{\mathbf{I}}_Q$ | Normalized projected Fisher information used when scale-invariant comparison of directional sensitivity or anisotropy is desired. | Eq. (12) |
| $\varepsilon$ | Tolerance threshold defining task-dependent information equivalence between sensing systems over a specified quotient-space region $\Omega$. | Eq. (10) |
| **Stage 4: Implications for Observable World Models** | | |
| $\mathcal{W}([\theta])$ | Observable world model defined on equivalence classes in $\Theta/\mathcal{G}$, rather than on redundant ambient parameters. | Section V |
| $b_t([\theta])$ | Belief over observable equivalence classes, used for quotient-space inference, prediction, and action refinement. | Section V |
| $\mathbf{I}_Q^{-1}$ | Local uncertainty structure associated with quotient-space distinguishability, used to organize uncertainty in the observable world model. | Section V |
| **Operational Diagnostics** | | |
| $\mathcal{D}(\mathcal{T};\theta)$ | Trajectory Diversity Index (TDI), commonly defined through the weakest quotient-space Fisher direction, e.g., $\lambda_{\min}(\mathbf{I}_Q)$. | Eq. (17) |
| $\lambda_{\min}(\mathbf{I}_Q)$ | Smallest eigenvalue of the projected Fisher information, indicating weakest-direction excitation on the quotient space. | Eq. (17) |
| $\kappa$ | Geometry–material coupling coefficient, measuring interaction between geometric and material sensitivity subspaces. | Section VI-E |
| $\rho$ | Jacobian correlation metric, indicating redundancy across sensing configurations or trajectory states. | Section VI-E |
| RSR | Residual-to-signal ratio, quantifying mismatch between the physical observation model and measured data. | Section VI-E |

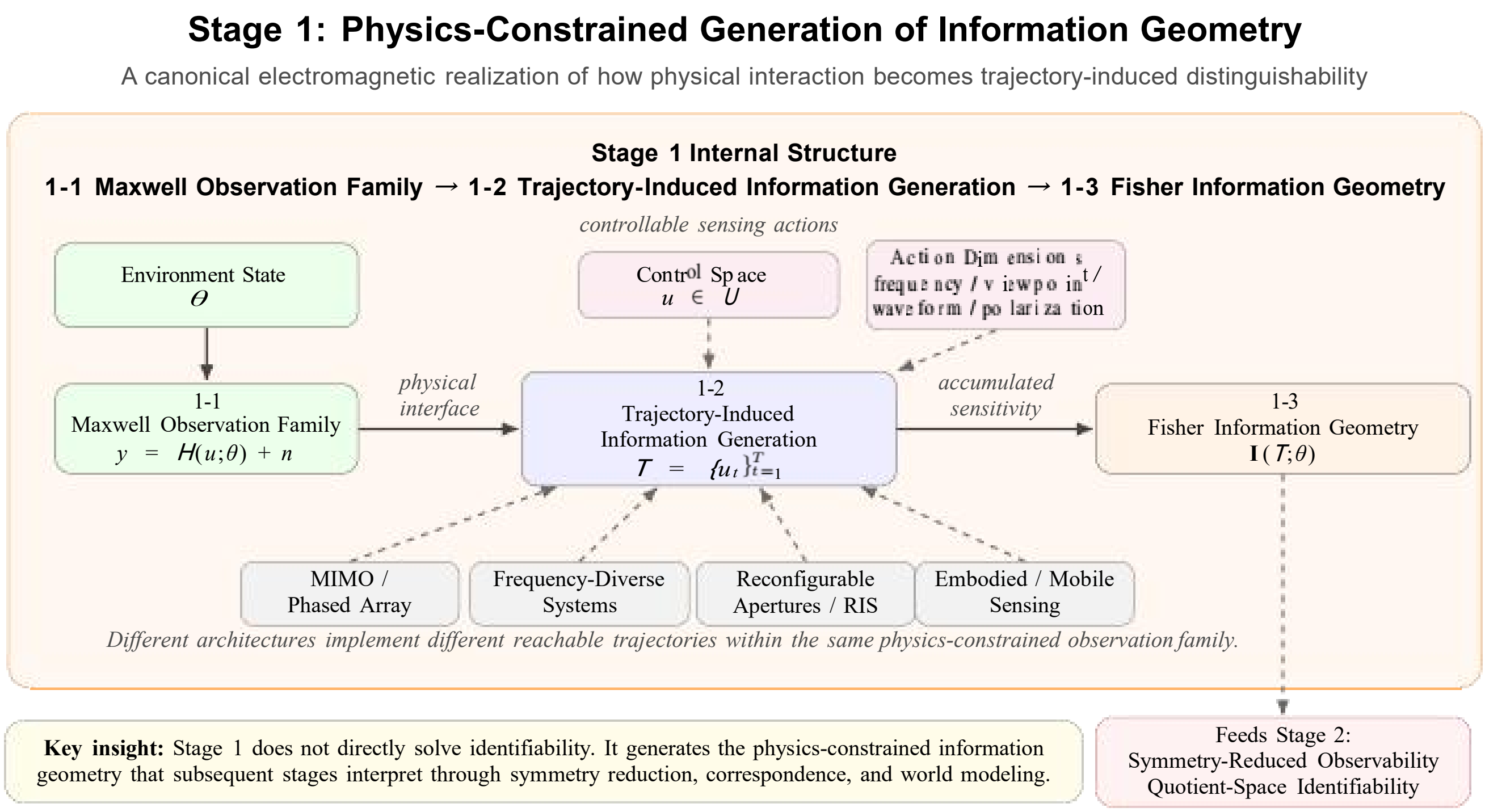

Fig. 4. **Stage 1 as physics-constrained generation of information geometry.** This figure refines the first stage of the observability-engineering framework. Electromagnetic sensing begins with an environment state $\theta$ and a Maxwell-consistent observation family $y = H(u;\theta) + n$. Controllable sensing actions $u \in U$ generate sensing trajectories $\mathcal{T}$, which accumulate local sensitivity and induce Fisher information geometry $\mathbf{I}(\mathcal{T};\theta)$. Quotient-space identifiability is shown only as the next stage, not as part of the Stage 1 mechanism.

In practice, the exact parameter $\theta$ is unknown, and Fisher information must therefore be evaluated under uncertainty [7, 36]. Several approximations are commonly used in engineering systems. First, a nominal or reference parameter $\theta_0$ can be used, yielding a local design criterion commonly adopted in adaptive sensing [59, 40]. Second, Bayesian experimental design can replace $\mathbf{I}(\mathcal{T};\theta)$ with its expectation under a prior distribution $p(\theta)$ [42, 95]. Third, in closed-loop systems, $\theta$ can be replaced by online estimates $\hat{\theta}_t$, enabling recursive adaptation of sensing interactions [81, 96].

The quotient representation is used here primarily as a local design and analysis tool. For the quotient $\Theta/G$ to admit a smooth local structure, we assume that the symmetry group $G$ acts freely and properly on the domain of interest [72, 93, 97]. Under these conditions, projection onto quotient-identifiable directions becomes locally well defined [98]. These assumptions may fail at singular configurations, such as coincident targets or degenerate scattering responses. In such cases, the quotient may exhibit singular or stratified structure, and the projected Fisher geometry should be interpreted as a local structural approximation rather than a globally smooth manifold construction [93].

## *B. Stage 2- 1: Quotient Representation of Observable States*

Stage 2-1 establishes the quotient representation of observable states. Let the environment state be parameterized by $\theta \in \Theta$. In many sensing systems, transformations exist that leave the observation invariant. We assume a group action of $G$ on $\Theta$ such that

$$p(y|u;\theta) = p(y|u;g\cdot\theta), \quad \forall u \in U,\ g \in G, \tag{6}$$

i.e., the observation distribution is invariant under the symmetry group $G$ [71, 92, 72].

If the invariance holds for all admissible sensing actions $u \in U$, then no sensing interaction can distinguish $\theta$ from $g\cdot\theta$. Two parameter states are observationally equivalent if they differ by such a transformation. The equivalence class

$$[\theta] = \{g\cdot\theta : g \in G\}$$

defines the observable state, and the quotient space

$$\Theta_{\text{obs}} = \Theta/G$$

represents what can be distinguished in principle [72, 99].

Different sensing scenarios admit different symmetry groups and therefore different quotient structures, as summarized in Table V.

This quotient representation resolves ambiguity by explicitly removing symmetry-induced degeneracy, thereby isolating the component of indistinguishability that can be addressed through sensing design. The quotient representation is therefore not merely a mathematical convenience, but a necessary structure for separating intrinsic ambiguity from design-dependent observability [93, 98].

Identifiability is therefore not a property of $\theta$, but of its equivalence class $[\theta]$. This representation removes redundancy associated with symmetry and defines the intrinsic state space of the sensing system [71].

TABLE V
SUMMARY OF SYMMETRY-INDUCED QUOTIENT SPACES FOR COMMON SENSING SCENARIOS

| **Sensing Scenario** | **Full Parameter Space $\Theta$** | **Symmetry Group $\mathcal{G}$** | **Observable Quotient $\Theta/\mathcal{G}$** |
|---|---|---|---|
| Monostatic radar (2D) | $(x, y, \sigma)$ | $SO(2)$ rotation | $(R, \sigma)$ range + reflectivity |
| Polarimetric radar | $\mathbf{S} \in \mathbb{C}^{2\times 2}$ | $U(1)$ absolute phase | $[\mathbf{S}]$ relative phases + magnitudes |
| Bistatic radar | $(x, y, \sigma)$ | Translation invariance | Range-difference + reflectivity |
| SAR/ISAR | $(x, y, \dot{x}, \dot{y}, \sigma)$ | Time shift | Motion-compensated position + reflectivity |

*1) Example 1: Rotational Symmetry in Monostatic Radar:* Consider the monostatic radar system illustrated in Fig. 3. A target at position $(x, y)$ with reflectivity $\sigma$ is observed through its range

$$R = \sqrt{x^2 + y^2},$$

while the azimuth angle $\phi = \arctan(y/x)$ is not measured. The observation therefore depends only on $(R, \sigma)$ [3, 4].

The rotational symmetry of this system is shown in Fig. 5, where all targets lying on the same circle of radius $R$ are observationally equivalent. This invariance corresponds to the rotation group $\mathcal{G} = SO(2)$, under which the symmetry orbit $[\theta]$ is a circle in the $(x, y)$ plane, and the angle $\phi$ is fundamentally unobservable [93, 97].

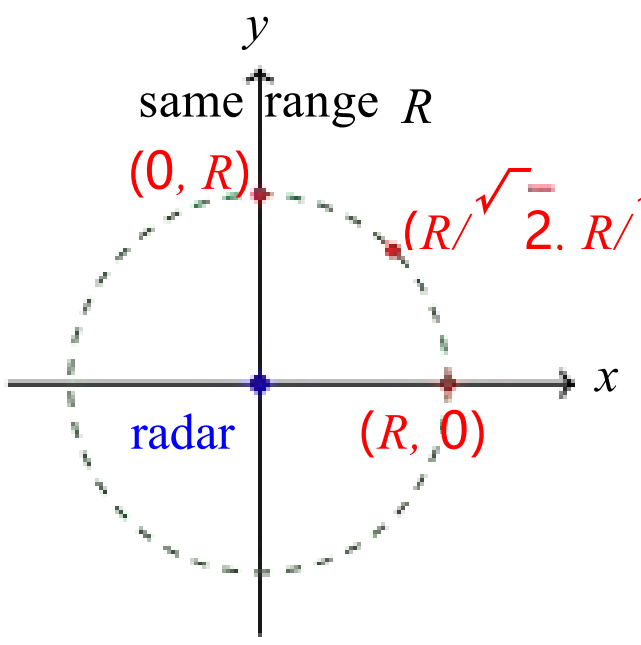

Fig. 5. Rotational symmetry in monostatic radar. All targets on the same range circle produce identical measurements. The observable state therefore reduces to the equivalence class $(R, \sigma)$ rather than the ambient coordinate $(x, y, \sigma)$.

A key insight is that the observable state is not the absolute position $(x, y)$, but the equivalence class $(R,\sigma)$. This distinction between what is measured and what can be inferred forms the central theme of quotient-space identifiability.

*2) Example 2: Polarization Ambiguity and Quotient Space:* The polarization example illustrates a symmetry-induced ambiguity in scattering-coefficient space rather than physical position space [100, 101].

A polarimetric radar target is characterized by the Sinclair matrix $\mathbf{S} \in \mathbb{C}^{2\times 2}$, and the received field for transmitted polarization $\mathbf{E}_t$ is $\mathbf{E}_r = \mathbf{S}\mathbf{E}_t$ [102, 103].

For measurement models that do not preserve an absolute phase reference, the observable statistics are invariant under a global phase rotation:

$$p(y \mid u; \mathbf{S}) = p(y \mid u; e^{j\phi_0}\mathbf{S}), \qquad \forall \phi_0 \in [0, 2\pi). \tag{7}$$

The equivalence class of a Sinclair matrix is

$$[\mathbf{S}] = \{e^{j\phi_0}\mathbf{S} : \phi_0 \in [0, 2\pi)\}.$$

The quotient space is therefore

$$\Theta_{\text{obs}} = \mathbb{C}^{2\times 2}/U(1),$$

which removes the globally unobservable phase degree of freedom [100, 104].

Many radar systems implicitly operate on this quotient space by forming polarimetric target decompositions that are invariant to absolute phase [100, 104]. These decompositions can be interpreted as constructing coordinates on $\Theta/\mathcal{G}$, thereby focusing estimation effort on observable quantities [102].

## *C. Stage 2- 2: Projected Fisher Information on the Quotient*

Stage 2-2 introduces projected Fisher information on the quotient-identifiable subspace. At $\theta \in \Theta$, the symmetry orbit

$$\mathcal{O}_\theta = \{g \cdot \theta : g \in \mathcal{G}\}$$

has tangent space $T_\theta \mathcal{O}_\theta$. Directions tangent to the orbit correspond to perturbations that leave the observation invariant to first order [72, 98].

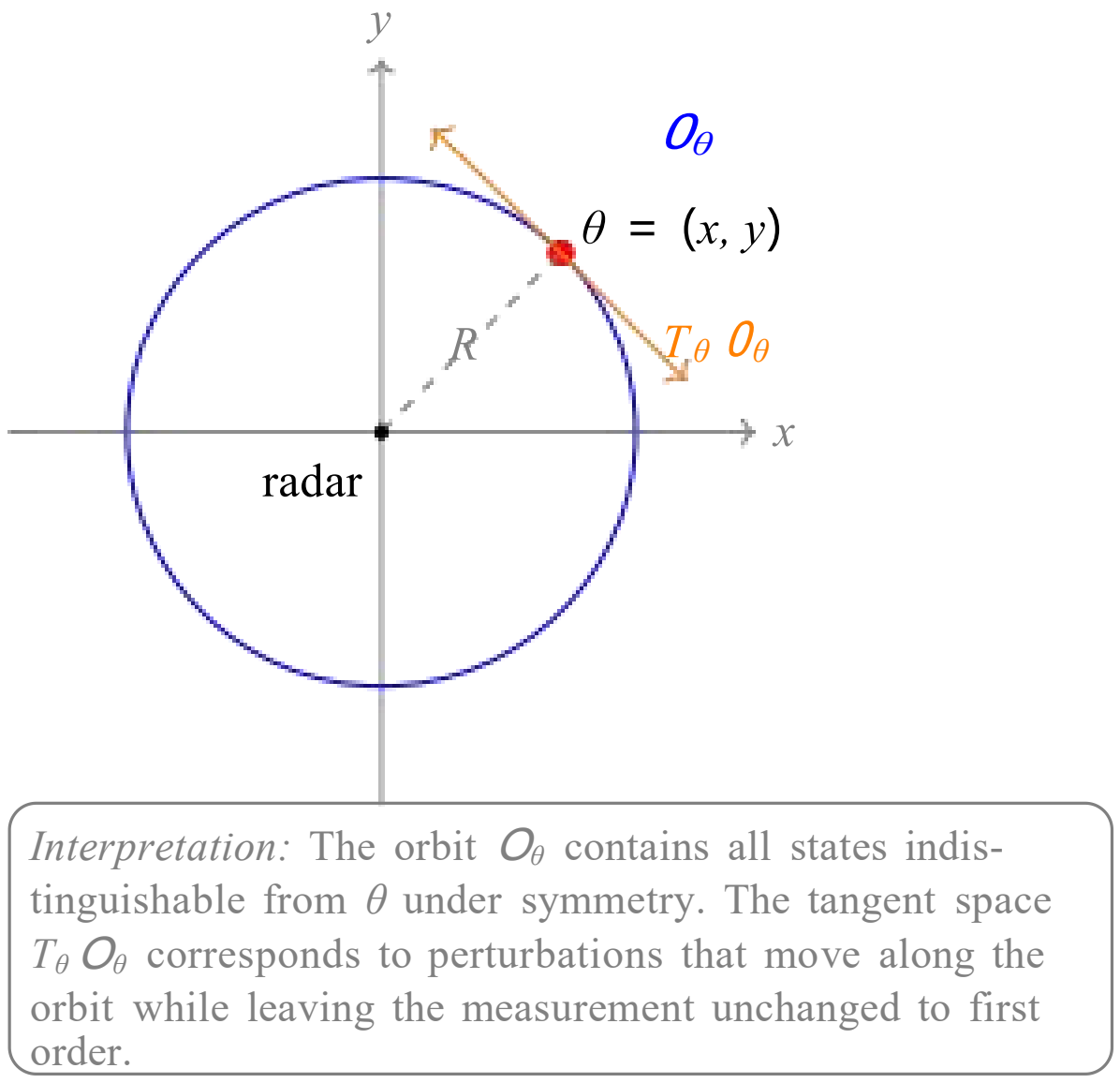

Fig. 6. Orbit geometry under rotational symmetry in monostatic radar. The quotient-identifiable direction is transverse to the symmetry orbit, while tangent directions remain fundamentally unobservable.

To quantify distinguishability after symmetry reduction, let $\mathbf{I}(\theta)$ denote the Fisher information matrix in the ambient parameter space. Let $\mathbf{P}_\perp(\theta)$ denote the projection onto a complement of $T_\theta \mathcal{O}_\theta$, defined with respect to a chosen reference inner product [72]. The quotient-relevant Fisher information is then given by

$$\mathbf{I}_Q(\theta) = \mathbf{P}_\perp^T(\theta)\, \mathbf{I}(\theta)\, \mathbf{P}_\perp(\theta), \tag{8}$$

which captures distinguishability restricted to quotient-identifiable directions [43, 105].

For a sensing trajectory $\mathcal{T} = \{u_t\}_{t=1}^{T}$, the trajectory-accumulated projected Fisher information is

$$\mathbf{I}_Q(\mathcal{T};\theta) = \sum_{t=1}^{T} \mathbf{I}_Q(u_t;\theta), \tag{9}$$

under standard independence assumptions across measurements. In correlated or adaptive sensing settings, this additive form may no longer hold exactly, and $\mathbf{I}_Q(\mathcal{T};\theta)$ should instead be interpreted as a trajectory-level information functional on the quotient space.

The choice of complement is not unique; however, different complements correspond to different coordinate representations of the quotient-identifiable subspace. The design-relevant quantity is therefore the induced Fisher geometry transverse to the symmetry orbit.

A key insight is that only perturbations orthogonal to the symmetry orbit contribute to distinguishability and identifiability, while tangent directions remain fundamentally unobservable.

In the monostatic radar example, the orbit $\mathcal{O}_\theta$ is the circle of constant range. The tangent space $T_\theta\mathcal{O}_\theta$ corresponds to angular perturbations that do not affect the measurement. These directions lie in the null space of the Jacobian $\mathbf{J}(\theta)$. Since the Fisher information satisfies $\mathbf{I} = \mathbf{J}^T\mathbf{\Sigma}^{-1}\mathbf{J}$ under standard regularity conditions [7], such directions contribute no Fisher information and are therefore unobservable. By contrast, perturbations orthogonal to the orbit change the range and are observable. More generally, only directions orthogonal to the symmetry orbit contribute to identifiability [71].

For the polarization example, the tangent space $T_\theta\mathcal{O}_\theta$ corresponds to the direction of global phase rotation in the real parameter space of $\mathbf{S}$. This direction lies in the null space of the Jacobian for any transmitted polarization, confirming that absolute phase is fundamentally unobservable [100, 101].

Local identifiability on the quotient space is determined by the projected Fisher information $\mathbf{I}_Q(\theta)$ on the quotient-identifiable subspace [71, 92]. When $\mathbf{I}_Q(\theta)$ is positive definite, the corresponding equivalence class becomes locally distinguishable in principle. By contrast, if $\mathbf{I}_Q$ is singular, there exists a direction orthogonal to the symmetry orbit that remains indistinguishable, indicating a fundamental ambiguity in the observable structure. This condition is invariant under reparameterization of $\Theta/\mathcal{G}$, ensuring that identifiability is a geometric property of the quotient manifold rather than a coordinate artifact.

The quotient-space formulation separates intrinsic symmetry-induced degeneracy from design-dependent excitation effects. Symmetry-induced degeneracy is removed through quotienting and reveals what cannot be observed regardless of sensing strategy [72, 71]. By contrast, remaining rank deficiency or poor conditioning of $\mathbf{I}_Q$ reflects insufficient excitation of observable directions and may be mitigated through sensing design [59, 81, 96].

This distinction has important implications for sensing-system design. The quotient space determines the observable structure of the sensing problem, while controllable sensing interactions determine whether these observable directions are sufficiently excited [32, 28, 29]. From this perspective, sensing design can be interpreted as shaping the spectrum and conditioning of $\mathbf{I}_Q$ through controllable sensing actions.

While projected Fisher information on the quotient space resolves symmetry-induced degeneracy, it retains several important limitations [82, 83]. First, the resulting information geometry remains inherently local and may not capture global ambiguities arising from strong nonlinearities or disconnected solution structures. Second, the formulation relies on regularity assumptions, including smoothness and locally Gaussian noise behavior, which may not hold universally across all sensing regimes. Finally, the projection operator $\mathbf{P}_\perp$ generally depends on the underlying parameter state $\theta$, which may itself be imperfectly known in practical sensing systems.

Nevertheless, by separating intrinsic symmetry-induced degeneracy from design-dependent excitation effects, the quotient-space formulation establishes a principled geometric foundation for observability analysis, information generation, and controllable sensing interaction within a unified framework.

## IV. Stage 3: Cross-System Information Equivalence

Stages 1 and 2 established the information-theoretic domain on which sensing systems should be evaluated. Stage 1 showed how controllable sensing actions generate Fisher information geometry through physics-constrained observation families [43, 7], while Stage 2 identified the observable quotient space $\Theta/\mathcal{G}$ on which distinguishability is meaningful [71, 72]. Stage 3 now asks a different engineering question: when should two physically different sensing systems be regarded as functionally equivalent for a given task?

In the present framework, this question is answered at the level of quotient-space information structure. Two architectures should not be compared only by hardware descriptors such as aperture size, number of antennas, bandwidth, transmit power, or waveform class. Rather, they should be compared by how they distinguish observable states on the task-relevant region of $\Theta/\mathcal{G}$ [41, 40]. Thus, cross-system equivalence is defined by the information structures that sensing systems induce on the same observable quotient space, rather than by similarity of physical implementation.

This section formalizes this idea through regional information correspondence. Given two sensing architectures $\mathcal{A}$ and $\mathcal{B}$, the goal is to compare their induced quotient-space Fisher structures over a task region $\Omega \subseteq \Theta/\mathcal{G}$. This provides a basis for hardware-agnostic benchmarking, cross-platform transfer, and resource substitution across spatial, spectral, waveform, polarization, and trajectory dimensions.

### A. Motivation: Beyond Hardware Comparison

Traditional sensing system design compares architectures through hardware descriptors: aperture size, number of antennas, transmit power, carrier frequency, and waveform class. While these parameters are physically meaningful, they do not directly capture functional sensing capability—namely, *how well a system distinguishes between different environment states* [3, 4].

Consider the following observations:

- A monostatic radar flying a long synthetic aperture may achieve angular resolution comparable to a large phased array radar, despite having far fewer antennas [15, 17].
- A frequency-diverse radar with a single antenna may achieve range resolution comparable to a wideband pulsed radar [57, 56].
- A small, mobile platform executing a diverse trajectory may outperform a larger, static array with superior hardware but limited diversity [28, 32].

These examples suggest a central principle: *Sensing architectures should be compared through the information structures they induce on the observable quotient space, rather than through hardware parameters alone.* This perspective aligns with information-theoretic approaches to sensing system design [106, 74].

### B. Information Correspondence on the Quotient Space

Stage 2 established that sensing performance should be evaluated on the observable quotient space $\Theta/G$, rather than on the redundant ambient parameterization [71, 72]. Stage 3 uses this quotient-space information structure to compare different sensing architectures [43, 41].

For a sensing architecture $A$ operating over a trajectory $T_A$, let $\mathbf{I}_Q^A(T_A;\theta)$ denote the projected Fisher information induced on $\Theta/G$ [7, 6]. This matrix summarizes how architecture $A$, together with its trajectory, locally distinguishes observable states near $\theta$ [43, 105]. Two architectures should therefore be compared not by their hardware parameters alone, but by the information structures they induce on the same observable quotient space [41, 40]. Informally, architectures $A$ and $B$ are *information-equivalent* over a task region $\Omega \subseteq \Theta/G$ if their induced quotient-space information structures are sufficiently close over $\Omega$:

$$D_{\text{corr}}(A, B; \Omega) \leq \varepsilon. \tag{10}$$

Here, $D_{\text{corr}}$ denotes a task-dependent discrepancy between information structures [42, 74], and $\varepsilon$ is an application-dependent tolerance.

> *Plain-language interpretation:* Two sensing systems may look different physically, but they can be functionally similar for a given task if they generate comparable information about the same observable quantities [15, 17, 34].

A regional notion of information correspondence can be introduced as follows. Let $\Omega \subseteq \Theta/G$ denote a task-relevant region, and let $\mu$ be a measure (e.g., uniform, prior-based, or task-weighted) over $\Omega$ [40, 42]. The information-correspondence discrepancy can then be defined as

$$D_{\text{corr}}(\mathcal{A},\mathcal{B};\Omega) = \int_{\Omega} d\left(\mathbf{I}_{\mathcal{Q}}^{\mathcal{A}}(\theta),\mathbf{I}_{\mathcal{Q}}^{\mathcal{B}}(\theta)\right)\, d\mu(\theta), \tag{11}$$

where $d(\cdot,\cdot)$ is a discrepancy measure between positive-semidefinite information matrices [41, 43]. In its simplest form, $d(\cdot,\cdot)$ may compare the spectral structure (e.g., eigenvalues) or the task-weighted estimation accuracy via the Cramér–Rao bound [7, 6].

When absolute estimation accuracy matters (e.g., SNR, power, or range requirements), the raw Fisher information matrices should be compared directly. When the goal is to compare directional sensitivity patterns independent of absolute scale, a normalized representation can be used [43]:

$$\tilde{\mathbf{I}}_{\mathcal{Q}}(\mathcal{T};\theta) = \frac{\mathbf{I}_{\mathcal{Q}}(\mathcal{T};\theta)}{\operatorname{tr}(\mathbf{I}_{\mathcal{Q}}(\mathcal{T};\theta)) + \delta}, \tag{12}$$

where $\delta > 0$ is a small regularization constant. This normalized form emphasizes anisotropy and the distribution of information across observable directions, rather than absolute information magnitude [105].

Different engineering tasks also require different notions of information correspondence. For applications where conditioning and weakest-direction excitation matter, comparing the eigenvalues of $\mathbf{I}_Q$ is appropriate. When different systems excite different subspaces of $\Theta/G$, comparing dominant eigenspaces or null spaces becomes relevant. For task-specific performance, a weighted Cramér–Rao bound can compare expected estimation accuracy only on observable quantities of interest [42, 41]. Table VI summarizes these comparison modes.

### C. Modes of Comparison and Practical Interpretation

Different engineering tasks require different notions of information correspondence. Table VI summarizes five common comparison modes. The appropriate choice depends on whether the task requires preserving absolute estimation accuracy, directional anisotropy, weakest-direction excitation, subspace alignment, or downstream task cost.

For example, absolute information should be compared when SNR and range performance matter. Normalized anisotropy is useful when the goal is to compare the shape of the information ellipsoid independent of scale. Spectral and subspace comparisons are useful when one wants to diagnose which observable directions are excited. For most engineering design problems, however, the most direct criterion is task-weighted cost, because it measures the consequence of information differences on the quantities that matter for the task.

Among the comparison modes in Table VI, the task-weighted criterion is often the most directly useful for engineering design. Let $\mathbf{W}(\theta)$ denote a task-weighting matrix that specifies which observable directions matter for the application. A local CRB-based discrepancy between architectures $A$ and $B$ can be written as

$$D_{\text{task}}(\mathcal{A},\mathcal{B}) = \left| \int_{\Omega} \operatorname{tr}\left[\mathbf{W}(\theta)\left((\mathbf{I}_{\mathcal{Q}}^{\mathcal{A}}(\theta) + \eta\mathbf{I})^{-1} - (\mathbf{I}_{\mathcal{Q}}^{\mathcal{B}}(\theta) + \eta\mathbf{I})^{-1}\right)\right] d\mu(\theta) \right|, \tag{13}$$

where $\eta > 0$ is a small regularization constant.

This criterion is useful because it compares systems through the estimation consequences of their information structures,

TABLE VI
MODES OF CROSS-SYSTEM INFORMATION COMPARISON ON THE OBSERVABLE QUOTIENT SPACE. THE TABLE SERVES AS A PRACTICAL GUIDE FOR SELECTING AN APPROPRIATE COMPARISON CRITERION DEPENDING ON THE ENGINEERING OBJECTIVE.

| Comparison Mode | What is Compared | When It Is Useful |
|---|---|---|
| Absolute information | R aw $\mathbf{I}_Q$ | When SNR, power, range, or estimation accuracy must be preserved. |
| Normalized anisotropy | $\mathbf{I}_Q$ | When comparing directional sensitivity patterns independent of scale. |
| Spectral structure | Eigenvalues of $\mathbf{I}_Q$ | When conditioning and weakest-direction excitation matter. |
| Subspace structure | Dominant eigenspaces or null spaces | When systems excite different observable directions. |
| Task-weighted cost | Weighted inverse-Fisher or CRB-based criteria | When only certain observable quantities matter for the task. |

rather than through the information matrices alone. It also makes explicit that information equivalence is task-dependent: two systems may be equivalent for range estimation but not for material classification or velocity estimation.

### D. Illustrative Examples and Design Implications

The information-correspondence perspective naturally explains several known phenomena in sensing-system design [41, 43]. In each case, the relevant comparison is not whether two systems share the same hardware implementation, but whether they induce comparable information structures on the task-relevant region of the observable quotient space [71, 72].

*1) Example 1: Moving monostatic vs. phased array.:* A moving monostatic radar that flies along a trajectory spanning an angular range $\Delta\psi$ can induce an information geometry on $\Theta/G$ similar to that of a phased-array radar with angular coverage $\Delta\psi$ for 2D target localization [15, 17, 3]. The equivalence holds only when the synthetic aperture, angular coverage, SNR, and observation time are matched [12, 4]. *The equivalence is not architectural; it is task- and trajectory-dependent.* This observation connects to the concept of synthetic aperture radar (SAR), where platform motion substitutes for physical spatial aperture [18, 16, 17].

*2) Example 2: Full MIMO vs. TDM-MIMO virtual aperture.:* Consider two radar architectures designed to form a comparable virtual aperture. Architecture $A$ uses simultaneous MIMO transmission and receives multiple transmit–receive channel responses in parallel [34, 76, 107]. Architecture $B$ uses time-division multiplexed MIMO (TDM-MIMO), activating transmitters sequentially over a coherent burst [54, 55]. Although the hardware and sampling process differ, both systems may induce similar quotient-space information structures for angle–range localization if their effective virtual aperture, SNR, waveform bandwidth, and motion compensation are matched [56, 57].

In the notation of Stage 3, the relevant question is not whether the two systems have the same number of RF chains or transmitters, but whether

$$\mathbf{I}_Q^A(T_A;\theta) \approx \mathbf{I}_Q^B(T_B;\theta)$$

over the task region $\Omega \subseteq \Theta/G$ [7, 6]. When the scene remains coherent over the TDM burst and motion-induced phase errors are compensated, TDM-MIMO can approximate the information geometry of a simultaneous MIMO system on the task-relevant quotient directions [26, 27]. When coherence is violated or Doppler coupling is not compensated, the equivalence breaks down even if the nominal virtual aperture is the same [108, 4]. This example illustrates that information equivalence is conditional: functional sensing capability is determined by the induced quotient-space information structure, not by hardware similarity alone [41, 43, 71].

*3) Example 3: Frequency diversity vs. spatial diversity.:* Frequency diversity and spatial diversity may excite different but partially overlapping observable directions on the quotient space [56, 19, 109]. Their equivalence should therefore be evaluated through $D_{\mathrm{corr}}(A, B; \Omega)$ over the task region, rather than assumed through a universal closed-form mapping [40, 41]. In practice, one architecture may be preferred over the other based on constraints such as spectral availability, platform size, hardware complexity, or coherence time [57, 34, 53, 54].

*4) Example 4: Polarization diversity across platforms.:* A polarimetric radar with a single antenna but sequentially transmitting multiple polarization states may induce a polarization-diversity structure on the quotient space comparable to a multi-antenna system with simultaneous polarimetric measurements, provided that the sequential measurements remain coherent or can be registered within the task model [101, 100, 104]. In this case, information correspondence requires the accumulated projected Fisher matrices after sequential polarization probing to match those obtained from simultaneous polarimetric measurements over the task region [105, 43]. This example illustrates that the same induced information geometry can arise from fundamentally different hardware implementations [102, 103, 100].

The information equivalence perspective enables several design and analysis capabilities [40, 41]:

- **Cross-platform transfer:** A sensing controller or inference module developed for architecture $A$ can be transferred to architecture $B$ if the two are information-equivalent over the task region. This is particularly valuable when $A$ is a high-fidelity (e.g., expensive, large-scale) system and $B$ is a low-cost, resource-constrained platform [32, 33].
- **Resource equivalence:** Limited resources (bandwidth, time, number of antennas, power) can be allocated across different diversity dimensions (spatial, spectral, polarimetric, temporal) guided by maintaining information equivalence to a reference architecture [59, 96]. This provides a principled framework for trading resource types while preserving sensing capability.
- **Unified benchmarking:** Comparing sensing systems based on information correspondence rather than hardware specifications enables fair, task-relevant benchmark-ing across heterogeneous architectures [26, 27]. Two systems can be ranked not by which has higher power or more antennas, but by which induces better-conditioned information geometry on the observable quotient space for the task.

*5) Limitations and practical considerations:* Information equivalence does *not* imply that two systems are physically interchangeable in all respects. It does not guarantee equal robustness to calibration error, environmental mismatch, latency, field-of-view constraints, or nonlocal ambiguities. It only states that, over a specified task region and under the adopted local statistical model, the systems induce comparable distinguishability structures on the observable quotient space [36, 7].

Several caveats apply to the information equivalence framework [82]:

- **Local approximation:** Fisher information captures local distinguishability; global ambiguities may induce non-equivalence even when the Fisher geometry matches locally [43];
- **Noise model sensitivity:** The comparison assumes similar noise characteristics; systems with different noise profiles may not be equivalent even if their deterministic Fisher structures match [7];
- **Region dependence:** Equivalence over one region $\Omega$ does not imply equivalence over another; the task region must be specified [40];
- **Computational cost:** Evaluating $D_{\text{corr}}$ over a region may be expensive for high-dimensional quotient spaces.

When applying information equivalence in practice, we recommend:

1) Specify the task region $\Omega$ and the measure $\mu$ explicitly;
2) Choose the discrepancy measure $d(\cdot, \cdot)$ based on the task (spectral, geometric, or task-weighted);
3) Regularize Fisher matrices when near singular ($\delta, \eta > 0$);
4) Validate equivalence through simulation or field data when possible;
5) Do not claim substitutability beyond the conditions under which equivalence was established.

Despite these limitations, when applied within its regime of validity (local, Gaussian approximations, well-characterized quotient space, specified task region), information equivalence provides a principled and practical language for comparing and transferring sensing capabilities across architectures [43, 41].

## V. Stage 4: Implications for Observable World Models

Stages 1–3 established the structural basis of observability engineering. Stage 1 showed how controllable sensing actions generate Fisher information geometry through physics-constrained observation families [43, 7]. Stage 2 identified the observable quotient space $\Theta/G$, on which distinguishability and identifiability should be evaluated [71, 72]. Stage 3 showed that heterogeneous sensing architectures can be compared through the quotient-space information structures they induce [41, 40].

Stage 4 draws the representation-level implication of this progression. Once sensing becomes adaptive, an internal world model is no longer optional: the system must maintain a task state that supports sequential inference, action selection, prediction, and control. Such a state should not be defined over ambient physical parameters arbitrarily. If the sensing process distinguishes states only through their equivalence classes in $\Theta/G$, then the internal world model should be organized over the same observable structure.

> *Key insight.* If a sensing system cannot distinguish $\theta$ from $g \cdot \theta$ for $g \in G$, then a world model that treats these states as distinct contains representational degrees of freedom that cannot be resolved by any admissible sensing trajectory [71, 92].

The natural state variable for world modeling is therefore the equivalence class

$$[\theta] \in \Theta/G,$$

rather than the ambient parameter $\theta \in \Theta$. A world model in this framework is not a reconstruction of the most detailed possible scene description. It is a task-relevant organization of observable equivalence classes, together with the uncertainty, predictive structure, and decision-relevant relations needed for adaptive interaction [31, 110].

Fig. 7 summarizes this transition from ambient parameterization to quotient-space world-model construction. Symmetry reduction maps $\Theta$ to observable states $[\theta]$, belief updates and projected Fisher information organize uncertainty on $\Theta/G$, and the resulting world model guides future sensing actions. In this way, Stage 4 closes the observability-engineering loop: sensing trajectories generate information, symmetry reduction defines the observable state, and quotient-space uncertainty guides trajectory refinement.

### A. Observable World Models

Let $W$ denote an internal world model used for inference, planning, or control. In the quotient-space formulation, $W$ should be defined as a representation of observable states,

$$W : \Theta/G \to Z,$$

where $Z$ is a task-level representation space. Equivalently, if $\Phi : \Theta \to Z$ is written as a map from the ambient parameter space, it should satisfy the invariance condition

$$\Phi(\theta) = \Phi(g \cdot \theta), \qquad \forall g \in G. \tag{14}$$

This condition, which aligns with principles of equivariant representation learning [111, 112], ensures that symmetry-equivalent states are not represented as distinct world states.

An observable world model should contain three components [70, 43]:

- **Observable state variables:** coordinates or features defined on $\Theta/G$, not on redundant directions in $\Theta$;
- **Uncertainty geometry:** local uncertainty shaped by the projected Fisher information $\mathbf{I}_Q$;
- **Action dependence:** an update mechanism reflecting how future sensing trajectories change information on $\Theta/G$.

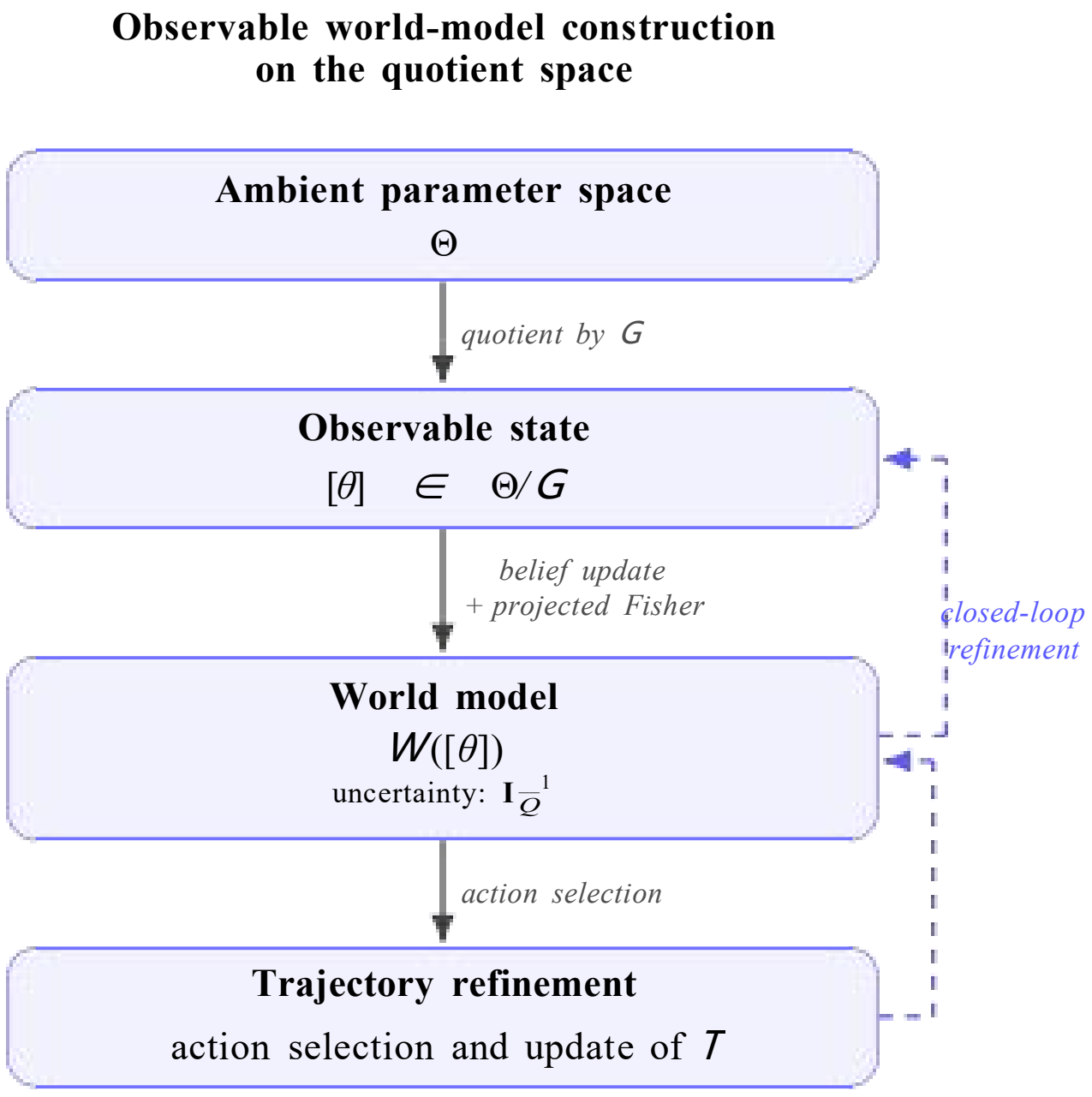

Fig. 7. Observable world-model construction on the quotient space. Ambient parameters $\theta \in \Theta$ are first reduced to observable equivalence classes $[\theta] \in \Theta/G$. Belief updates and projected Fisher information organize uncertainty on the quotient space, yielding a world model $W([\theta])$ whose uncertainty is structured by $\mathbf{I}_Q^{-1}$. The resulting world model guides action selection and trajectory refinement, closing the observability-engineering loop.

In this formulation, the role of the world model is not to encode everything about the environment, but to encode what the sensing system can distinguish and improve through action [28, 32].

## B. Belief Updating on the Quotient Space

A concrete way to express this idea is to maintain a belief over observable states. This follows the tradition of probabilistic robotics and active perception [31, 113]. Let $b_t([\theta])$ denote a belief distribution on $\Theta/G$. Given a sensing action $u_t$ and observation $y_t$, the belief update takes the form

$$b_t([\theta]) \propto b_{t\text{-}1}([\theta])p(y_t \mid u_t, [\theta]). \tag{15}$$

The likelihood $p(y_t \mid u_t, [\theta])$ is well defined because the observation distribution is invariant along each equivalence class, as stated in Eq. (6). Therefore, any representative $\theta \in [\theta]$ yields the same likelihood under admissible sensing actions [71]. This is precisely why the quotient space is an appropriate domain for world modeling.

Under local Gaussian approximations, the uncertainty of this belief is shaped by the inverse of the projected Fisher information [7, 6]. Directions with large eigenvalues of $\mathbf{I}_Q$ are well resolved, whereas directions with small eigenvalues remain uncertain [43]. Hence, the world model inherits not only the quotient-space state representation, but also the information geometry generated by sensing trajectories.

## C. Action Selection for World-Model Improvement

Because the world model is defined on $\Theta/G$, sensing actions should be selected to improve information on this quotient space. This connects to optimal experimental design and active perception [40, 114]. A standard local criterion is the expected information gain

$$G(\mathcal{T} \mid b_t) = \mathrm{E}_{[\theta]\sim b_t}\left[\log\det(\mathbf{I}_Q(\mathcal{T};[\theta]) + \epsilon\mathbf{I})\right], \tag{16}$$

where $\epsilon > 0$ is a numerical regularization constant [42, 95].

This criterion promotes volumetric information growth on the observable quotient space. It is complementary to the trajectory diversity index (TDI) introduced in Section VI: while the log-determinant criterion rewards total information volume, TDI diagnoses weakest-direction excitation [41]. Together, they support world-model improvement by encouraging both information accumulation and balanced excitation across observable directions [59, 96].

## D. Example: Monostatic Radar World Model

The monostatic radar example makes the implication concrete [3, 4]. Consider a point target observed by a monostatic radar transmitting a narrowband signal. In the ambient parameterization, the target may be represented as

$$\theta = (x, y, \sigma),$$

where $(x, y)$ denotes position and $\sigma$ denotes reflectivity. However, for a monostatic range-only observation, all points on the same range circle are indistinguishable, as illustrated in Fig. 5. The symmetry group is the rotation group $SO(2)$, and the observable quotient is represented by

$$[\theta] \equiv (R, \sigma), \qquad R = \sqrt{x^2 + y^2}.$$

A conventional world model that attempts to maintain $(x, y, \sigma)$ encodes an azimuthal degree of freedom that the sensing system cannot resolve. This can lead to unnecessary representational complexity and may cause downstream planning to pursue unobservable quantities [110, 115].

By contrast, an observable world model maintains $(R, \sigma)$ and assigns uncertainty only to directions that are meaningful under the observation model. If additional sensing actions, such as platform motion (synthetic aperture) or multi-view observations, break the rotational symmetry, the quotient structure changes accordingly, and the world model can be refined [15, 17].

This example illustrates the central point of Stage 4: world-model design should follow the observable structure induced by the sensing system, rather than assume that all ambient parameters are equally meaningful [31, 113].

## E. Implications for Representation Design

The quotient-space formulation gives three practical guidelines for designing world models in active sensing systems.

*1) Avoid representing unobservable coordinates as state variables:* If a degree of freedom lies along a symmetry orbit, it should not be treated as an independently recoverable state variable unless additional sensing actions break the symmetry. This principle aligns with the concept of minimal representations in robotics and equivariant learning [115, 111].

*2) Represent uncertainty in quotient-identifiable directions:* Uncertainty should be structured by $\mathbf{I}_Q$, not by the ambient Fisher matrix [43, 105]. This prevents the world model from confusing symmetry-induced redundancy with trajectory-induced uncertainty.

*3) Use sensing actions to refine the model:* A world model is not static. It is refined by selecting trajectories that increase information on weakly excited directions of $\Theta/\mathcal{G}$ [59, 40]. Thus, world modeling and trajectory design are coupled: the current world model guides sensing actions, and sensing actions reshape the information geometry of the world model [28, 32].

In summary, Stage 4 completes the observability-engineering chain. Stage 1 generates information geometry, Stage 2 removes symmetry-induced redundancy, Stage 3 compares induced information structures across systems, and Stage 4 defines how observable states should be internally represented. The resulting representation is not an arbitrary latent embedding, but a *physics-constrained, geometry-aware, action-refinable* world model defined on the observable quotient space.

> *A world model should not attempt to represent everything about the environment. It should represent what the sensing system can distinguish, how reliably it can be distinguished, and how future actions can improve it.*

## VI. Operational Observability: Design Principles and Observable Signatures

Sections II–V developed the four-stage structure of observability engineering. Stage 1 showed how physics-constrained sensing actions generate Fisher information geometry; Stage 2 identified the observable quotient space $\Theta/\mathcal{G}$ by removing symmetry-induced redundancy; Stage 3 compared heterogeneous sensing systems through their induced information structures; and Stage 4 organized observable quantities into task-relevant world representations.

This section translates the preceding structure into operational design principles and diagnostics for active sensing systems. In this view, observability is engineered by selecting, evaluating, and adapting sensing trajectories so that quotient-identifiable directions relevant to the task are sufficiently excited, robustly conditioned, and diagnosable through observable geometric signatures [59, 81, 40, 41].

This section synthesizes insights from Stages 1–4 into practical design tools. The emphasis is on how trajectory-level diagnostics, such as the trajectory diversity index (TDI), assess excitation quality on the observable quotient space, and how five complementary observable signatures reveal the health of the underlying information geometry. This perspective connects to classical ideas of persistent excitation, adaptive experiment design, and Fisher-based sensing criteria [59, 96, 40, 42, 95].

> *Core idea.* Observability is not directly measurable, but it reveals itself through consistent geometric signatures in Jacobians, Fisher information, and cross-configuration behavior. These signatures translate abstract information geometry into actionable engineering diagnostics.

### A. Closed-Loop Observability Engineering Workflow

The preceding stages provide a structural interpretation of observability; the engineering task is to turn this structure into an iterative design procedure. A closed-loop observability-engineering workflow can be summarized as follows:

1) Specify the observable quotient space $\Theta/\mathcal{G}$.
2) Evaluate trajectory-induced projected Fisher information $\mathbf{I}_Q(\mathcal{T};\theta)$.
3) Diagnose weak or redundant directions using TDI, conditioning, and Jacobian correlation.
4) Update the quotient-space belief or world model $\mathcal{W}([\theta])$.
5) Refine sensing actions to improve excitation of task-relevant observable directions.

This workflow turns observability from a descriptive property into an engineering procedure: identify what is observable, measure how well the current trajectory excites it, diagnose failure modes, update the internal quotient-space state, and refine future sensing actions accordingly.

### B. Quotient-Space Excitation and Trajectory Conditioning

*1) Action-induced excitation on the quotient:* Sensing actions do not remove the underlying physical symmetries; those symmetries have already been accounted for by the quotient representation [72, 93, 98]. What sensing trajectories control is the excitation of directions transverse to the symmetry orbits. In particular, a trajectory generates projected Fisher information on $\Theta/\mathcal{G}$, and this information determines how well the remaining quotient-identifiable directions are distinguished under local regularity assumptions [71, 92, 7, 6].

After symmetry-induced redundancy has been removed, the remaining projected Fisher information $\mathbf{I}_Q(\mathcal{T};\theta)$ describes how well a sensing trajectory excites observable directions [71, 7, 43]. Poor conditioning of $\mathbf{I}_Q$ indicates that some quotient-identifiable directions remain weakly excited, even though they are observable in principle [59, 96].

This distinction is central for design. The quotient space specifies what can be observed in principle, while the spectrum of $\mathbf{I}_Q(\mathcal{T};\theta)$ specifies how well a particular trajectory observes it in practice. Diagnostics based on this spectrum therefore quantify *trajectory-induced degeneracy* rather than intrinsic symmetry-induced non-identifiability [40, 41, 7].

*2) From identifiability to trajectory excitation:* While the quotient-space formulation clarifies what is identifiable in principle, it does not guarantee that a given sensing strategy sufficiently excites all identifiable directions [71, 92]. Even when $\Theta/\mathcal{G}$ is well defined, a particular trajectory $\mathcal{T}$ may yield a projected Fisher matrix that is rank-deficient or poorly conditioned [59, 96, 81].

This leads to a central engineering implication: improving sensing performance is fundamentally a problem of *trajectory design* on the quotient space. Increasing measurement precision or hardware capability alone is insufficient; sensing actions must be orchestrated to ensure that all identifiable directions on $\Theta/G$ are sufficiently excited and well-conditioned in the induced information geometry [40, 41, 42, 95].

*Engineering question.* When does a sensing trajectory effectively excite all identifiable degrees of freedom on $\Theta/G$?

*3) Common excitation failure:* A common failure arises when sensing design emphasizes a single form of diversity, such as bandwidth, while neglecting others, such as spatial or angular variation [53, 54, 55]. High aggregate information, measured for example by trace or determinant of $\mathbf{I}_Q$, does not preclude rank deficiency or poor conditioning. Information concentrated in a subspace of $\Theta/G$ leads to effective unobservability along orthogonal observable directions [59, 40, 41].

### C. Trajectory Diversity Index (TDI): A Core Diagnostic

To assess how a sensing trajectory excites the observable directions on the quotient space, we examine the spectral properties of the projected Fisher information matrix $\mathbf{I}_Q(\mathcal{T};\theta)$. The smallest eigenvalue $\lambda_{\min}(\mathbf{I}_Q)$ plays a fundamental role, as it characterizes the existence of observable directions on $\Theta/G$ [116, 117].

*Proposition (Observability Failure Boundary).:* Let $\mathbf{I}_Q(\mathcal{T};\theta)$ denote the Fisher information on the observable quotient space. Then:

- If $\lambda_{\min}(\mathbf{I}_Q) = 0$, the Fisher information matrix is rank-deficient on $\Theta/G$, and the system is locally non-identifiable; in particular, no unbiased estimator can achieve finite variance along at least one observable direction [71];
- If $\lambda_{\min}(\mathbf{I}_Q) > 0$, the Fisher information matrix is full-rank on $\Theta/G$, and the system is locally identifiable, admitting consistent estimation under standard regularity conditions [7, 36].

Thus, $\lambda_{\min}(\mathbf{I}_Q)$ defines a sharp boundary separating structural non-identifiability from recoverable estimation [59].

*Remark (Why $\lambda_{\min}$ matters).:* Scalar criteria such as trace or log-determinant quantify average or volumetric information, but do not detect the presence of degenerate directions [41]. In contrast, $\lambda_{\min}$ directly characterizes the existence of unobservable directions, making it uniquely suited for diagnosing observability failure [40].

*Engineering diagnostic (TDI).:* For system design and interpretation, we define the Trajectory Diversity Index (TDI) as

$$D(\mathcal{T};\theta) = \lambda_{\min}(\mathbf{I}_Q(\mathcal{T};\theta)). \tag{17}$$

Here, $D$ serves as a convenient diagnostic interface for trajectory evaluation (e.g., maximizing excitation or detecting degeneracy), while the underlying structural characterization is given by $\lambda_{\min}(\mathbf{I}_Q)$ [7].

*Positioning relative to classical observability.:* The TDI should be interpreted as an operational proxy for trajectory-level observability under finite-sample and noisy conditions, rather than a replacement for classical observability definitions (control-theoretic or Fisher-based) [37, 38]. It does not address global identifiability, nor does it capture non-local ambiguities [71]. Its role is diagnostic: it indicates whether a given trajectory adequately excites the observable directions identified by the quotient space under local Gaussian approximations. When TDI is zero or near-zero, the trajectory is failing to excite certain observable directions—but a positive TDI does not guarantee global observability. It is not intended as a universal optimality criterion, but as a practical diagnostic for excitation adequacy under local modeling assumptions [59, 81].

*Remarks on the definition.:* The TDI depends on the unknown parameter $\theta$; in practice, it is evaluated at a nominal estimate or averaged over a region of interest [42]. It is also scale-dependent: increasing frequency or sensor count generally increases $D$. Normalized variants based on trace or log-determinant may be used when scale invariance is required [41].

While not invariant under arbitrary reparameterizations, the TDI remains meaningful when evaluated on the quotient space $\Theta/G$ with a physically consistent parameterization [43]. The TDI is introduced as a diagnostic tool rather than a formal optimality criterion; alternative measures such as condition number or E-optimality may be used depending on the application [40].

### D. Operational Regimes and Design Implications

The TDI provides a scalar diagnostic that maps directly to design actions [59, 96]:

- **Under-excited regime** ($D = 0$): Certain observable directions are not excited at all. The remedy is to introduce new forms of diversity (viewpoint, frequency, polarization) rather than increasing measurement count [53, 101];
- **Ill-conditioned regime** ($D \approx 0$): All observable directions are technically excited but unevenly so. The remedy is to rebalance the trajectory, improving coverage across all dimensions of $\Theta/G$ [40, 41];
- **Well-excited regime** ($D > 0$ sufficiently large): All observable directions are sufficiently excited. Further improvement requires increasing signal-to-noise ratio or measurement count [7].

These regimes are not binary; the transition between them depends on noise level, numerical stability thresholds, and task-specific estimation tolerances [59].

*Threshold selection in practice.:* The TDI $D = \lambda_{\min}(\mathbf{I}_Q)$ provides a scalar diagnostic, but its interpretation requires a threshold to distinguish "well-excited" from "ill-conditioned" regimes [59]. In practice, we recommend the following approach:

1) Evaluate the empirical distribution of $D$ via Monte Carlo simulation around a nominal estimate $\hat{\theta}$, accounting for uncertainty in $\theta$ and noise realizations [42];
2) Set a threshold $\tau = \alpha \cdot \sigma^2$, where $\sigma^2$ is the noise variance and $\alpha$ is chosen between 10 and 100 depending on required estimation accuracy and numerical stability of the Fisher inversion [7];
3) If $D < \tau$, the trajectory is considered under-excited or ill-conditioned; design adjustments (e.g., introducing new diversity dimensions) are recommended [40].

TABLE VII
OPERATIONAL REGIMES AND DESIGN IMPLICATIONS OF TRAJECTORY DIVERSITY

| Regime | TDI Condition | Implication | Design Action |
|---|---|---|---|
| Under-excited | $D = 0$ | Structural blind spots; missing directions | Introduce new sensing diversity (view, frequency, polarization) |
| Ill-conditioned | $D \approx 0$ | Unstable estimation; noise-sensitive | Rebalance trajectory; improve diversity distribution |
| Well-excited | $D > 0$ | Stable, balanced information accumulation | Maintain diversity; optimize efficiency |

These guidelines are heuristic; the optimal threshold is task-dependent and may be calibrated via domain-specific simulation or cross-validation [41]. For systems with well-characterized noise, a more rigorous approach is to set $\tau$ such that the corresponding Cramér–Rao bound satisfies the application's accuracy requirements [6].

## E. Observable Signatures of Information Geometry

While the TDI provides a single scalar diagnostic, a more complete picture emerges from five complementary signatures that reflect how information geometry manifests in practice [43, 105]. Figure 8 shows how these five indicators project onto a common information-geometric structure. These diagnostics are not ad hoc heuristics, but direct consequences of the quotient-space framework developed in Sections III–VI. Each diagnostic reflects a specific geometric property of the induced information structure on $\Theta/G$, and can be evaluated using standard estimation outputs [7, 6].

*1) Signature 1: Geometry–material decoupling:* In high-frequency regimes, geometric parameters predominantly affect phase, while material parameters affect amplitude [63, 51]. This separation induces a structured Jacobian that can be assessed through the coupling coefficient

$$\kappa = \frac{\|\mathbf{J}_{\text{geo}}^\top \mathbf{J}_{\text{mat}}\|_F}{\|\mathbf{J}_{\text{geo}}\|_F \|\mathbf{J}_{\text{mat}}\|_F},$$

where $\mathbf{J}_{\text{geo}}$ and $\mathbf{J}_{\text{mat}}$ denote Jacobian blocks [117]. Persistent coupling ($\kappa > 0.7$) may indicate insufficient diversity or model limitations; strong decoupling ($\kappa < 0.3$) is consistent with the geometric regime [63].

*2) Signature 2: Information growth:* Observability requires excitation of all directions in the quotient space. This is precisely what the TDI quantifies: $D \to 0$ indicates degeneracy, while $D > 0$ reflects growth [59]. More directly, excitation can be assessed through the minimum eigenvalue $\lambda_{\min}$ of the projected Fisher information $\mathbf{I}_Q$. Approximately linear growth with the number of configurations suggests consistent excitation, whereas stagnation indicates trajectory deficiencies [40, 41].

*3) Signature 3: Configuration diversity:* Redundant sensing configurations limit effective information gain [40]. Diversity can be quantified through the average pairwise correlation of normalized Jacobians [117]:

$$\rho = \frac{2}{N(N-1)} \sum_{i<j} \frac{|\mathbf{J}_i^\top \mathbf{J}_j|}{\|\mathbf{J}_i\| \|\mathbf{J}_j\|}.$$

High correlation ($\rho > 0.7$) indicates redundancy; low correlation ($\rho < 0.5$) suggests effective diversity. Increasing diversity often yields greater benefit than increasing measurement count alone, consistent with experimental design principles [40, 41].

*4) Signature 4: Cross-configuration agreement:* Different sensing ensembles should yield consistent interpretations on the observable quotient space [71]. This can be assessed by comparing estimates after alignment over the symmetry group. Let $\hat{\theta}_A$ and $\hat{\theta}_B$ be estimates from two ensembles. Their distance on the quotient space,

$$d([\hat{\theta}_A], [\hat{\theta}_B]) \leq \varepsilon,$$

should be consistent with the Cramér–Rao bound under local approximation for consistent systems [7, 6]. Persistent disagreement may indicate model mismatch, incorrect symmetry assumptions, or insufficient excitation [71].

*5) Signature 5: Physics–data consistency:* Fisher-based interpretation assumes that the forward model captures the dominant structure of the data [63, 70]. This can be assessed through the residual-to-signal ratio:

$$\text{RSR} = \frac{\|\mathbf{y} - \mathbf{h}(\hat{\theta})\|}{\|\mathbf{y}\|}.$$

Structured residuals (e.g., systematic patterns) or large RSR ($> 0.1$) indicate model mismatch, in which case information-geometric interpretations should be treated with caution [63].

TABLE VIII
DIAGNOSTIC THRESHOLDS FOR OBSERVABLE SIGNATURES

| Diagnostic | Good Signature | Red Flag |
|---|---|---|
| Geometry–material coupling | $\kappa < 0.3$ | $\kappa > 0.7$ |
| Information growth | $\lambda_{\min} \propto N$ | $\lambda_{\min}$ saturates |
| Configuration diversity | $\rho < 0.5$ | $\rho > 0.7$ |
| Cross-configuration agreement | $d <$ CRB | Systematic bias |
| Physics–data consistency | RSR $< 0.1$ | RSR $> 0.1$ |

These thresholds are representative operating guidelines and may vary with system scale, noise level, and modeling assumptions [59, 81].

## F. Failure Modes and Practical Diagnostics

When the observable signatures depart from their healthy regimes, specific failure modes can be diagnosed. The abstract identifies several common failure modes in sensing system design [71, 92, 59]. This subsection maps each failure mode to its geometric signature within the quotient-space framework and provides diagnostic indicators.

*1) Failure Mode 1: Blind directions due to unresolved symmetry.:* When the sensing system does not break a physical symmetry, certain parameter directions remain fundamentally unobservable regardless of trajectory design. In the monostatic radar example (Section III-B1), rotational symmetry renders the azimuth angle $\phi$ unobservable; in polarimetric radar (Section III-B2), the global phase ambiguity persists even under arbitrary trajectories [100, 101].

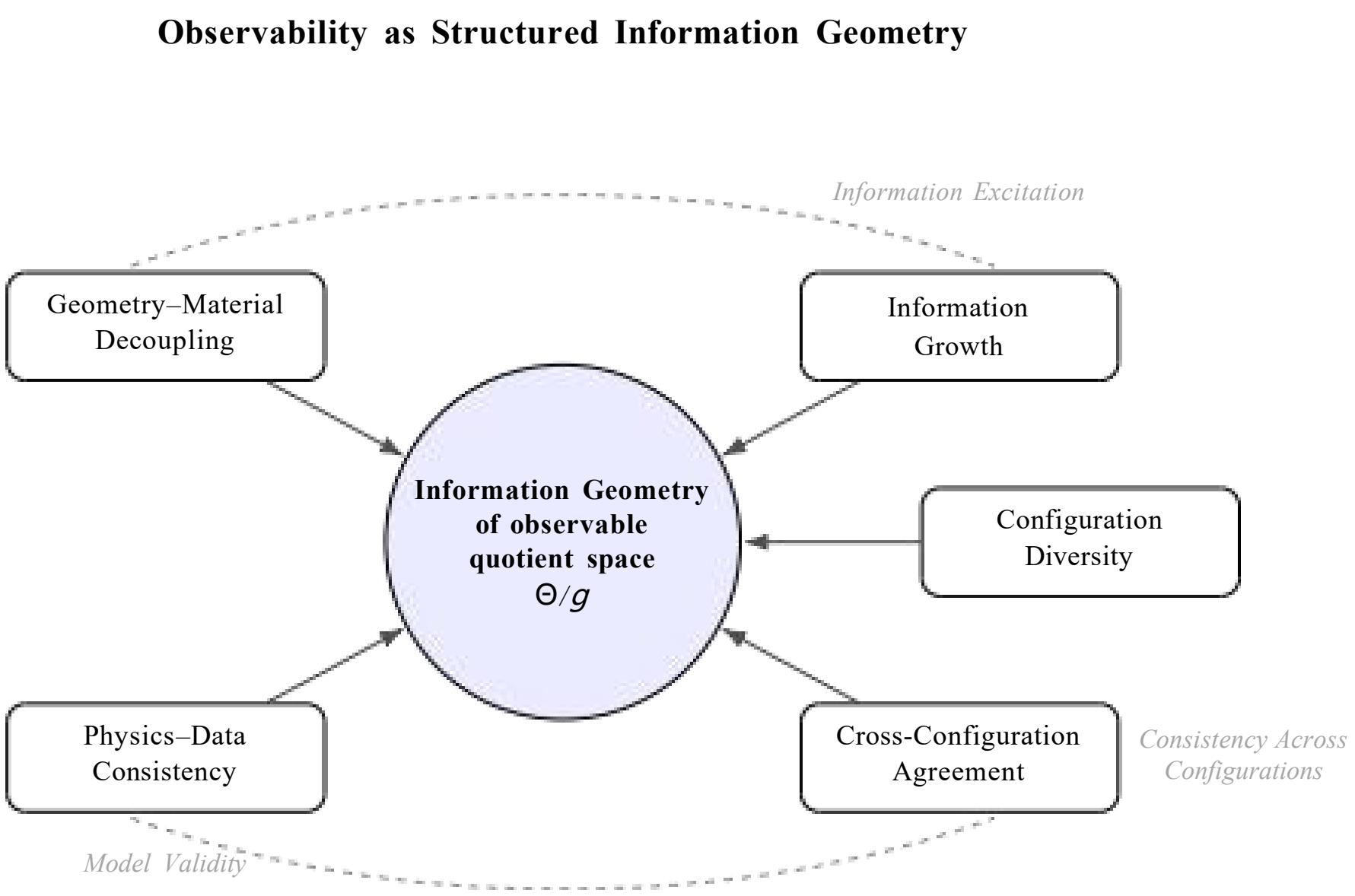

Fig. 8. **Unified interpretation of practical observability indicators.** The five indicators are different observable projections of the underlying information geometry on the symmetry-reduced space $\Theta/G$. Rather than forming an independent checklist, they reveal complementary aspects of a common structure: model validity (physics–data consistency), information excitation (growth and diversity), and cross-configuration consistency (agreement). Together, they offer an interpretable lens for assessing whether a sensing system generates meaningful and stable information.

*Signature:* The projected Fisher information $\mathbf{I}_Q$ is singular along the symmetry orbit direction, and the TDI is zero even with maximally diverse trajectories [71].

*Remedy:* Reformulate inference on the quotient space $\Theta/g$ (Section III) and, where possible, introduce symmetry-breaking diversity (e.g., multi-view configurations, relative phase references) [72, 101].

*2) Failure Mode 2: Ill-conditioned estimation from imbalanced excitation.:* Even when all directions are identifiable in principle, a trajectory that excites some directions much more strongly than others leads to poor conditioning [59, 96].

*Signature:* The condition number of $\mathbf{I}_Q$ is large ($\kappa > 10^3$), and the TDI is positive but small relative to noise variance [7, 117].

*Remedy:* Rebalance the trajectory by introducing diverse sensing actions (multiple polarization states, frequency steps) to improve excitation uniformity [40, 41].

*3) Failure Mode 3: False identifiability from insufficient diversity.:* A trajectory may appear to provide full-rank Fisher information when analyzed in the ambient parameter space, while fundamental ambiguities persist on the quotient space. This occurs when the analysis fails to account for symmetry-induced equivalence [71, 92].

*Signature:* $\mathbf{I}(\theta)$ (ambient) is full rank, but $\mathbf{I}_Q(\theta)$ (projected) is singular. The TDI computed on the ambient space overestimates observability.

*Remedy:* Always perform symmetry reduction before Fisher analysis. Compute $\mathbf{I}_Q$ using the projection $\mathbf{P}_\perp$ defined in Eq. (8) [72].

*4) Failure Mode 4: Physics–data mismatch.:* When the forward model $H(u;\theta)$ does not capture the dominant physical structure (e.g., strong multiple scattering, unmodeled coupling), the Fisher-based interpretation becomes unreliable [63, 51, 65].

*Signature:* Residual-to-signal ratio RSR $> 0.1$ or structured residuals (Signature 5) [70].

*Remedy:* Augment the model with data-driven residuals or switch to a full-wave forward model; treat Fisher-based diagnostics with caution [118, 119].

These failure modes are not mutually exclusive; a real system may exhibit multiple signatures simultaneously. The diagnostic framework in this section provides tools to isolate and identify each contribution.

### G. Integrated Design Workflow

The trajectory-centric interpretation of observability reframes sensing system design as a problem of *information shaping* on the quotient space rather than hardware optimization [59, 40]. The diagnostics developed in this section suggest a closed-loop workflow for sensing system design [40, 59]:

1) **Establish model credibility** through physics–data consistency (Signature 5);
2) **Assess excitation** via TDI and eigenvalue growth (Signatures 1–2);
3) **Check diversity** via Jacobian correlation (Signature 3);
4) **Validate consistency** via cross-configuration alignment (Signature 4);
5) **Iterate trajectory design** based on diagnosed failure modes.

*System-level insights.:* Several key insights emerge from this operational perspective:

- **Observability is trajectory-dependent:** The same hardware yields different observability under different sensing trajectories [28, 32];

TABLE IX
SUMMARY OF FAILURE MODES, SIGNATURES, AND REMEDIES

| Failure Mode | Signature | Remedy |
|---|---|---|
| Unresolved symmetry | $\mathbf{I}_Q$ singular, $\mathcal{D} = 0$ | Quotient reformulation + symmetry-breaking diversity |
| Imbalanced excitation | $\kappa(\mathbf{I}_Q) \gg 1$, $\mathcal{D} \approx 0$ | Trajectory rebalancing across diversity dimensions |
| False identifiability | $\mathbf{I}$ full rank, $\mathbf{I}_Q$ singular | Always compute $\mathbf{I}_Q$ via $\mathbf{P}_\perp$ |
| Physics–data mismatch | RSR $> 0.1$, structured residuals | Model augmentation or full-wave forward model |

- **Trajectory design can substitute for hardware:** Observability can often be significantly improved through trajectory design without modifying hardware. By introducing diversity in sensing actions—through motion, frequency variation, or waveform adaptation—the induced information geometry on $\Theta/\mathcal{G}$ can be reshaped to better excite otherwise weak directions [15, 17];
- **Architecture equivalence via information geometry:** Distinct systems may achieve comparable observability if they induce similar information geometry on the quotient space, even if their physical implementations differ [41];
- **Multi-domain diversity is essential:** Effective sensing design requires distributing resources across multiple dimensions, including spatial coverage, frequency diversity, and waveform design [53, 54]. Concentrating resources in a single domain often leads to imbalanced excitation, whereas coordinated multi-domain diversity promotes robust and stable inference on $\Theta/\mathcal{G}$ [101, 100];
- **Symmetry-breaking as design strategy:** Many limitations in sensing systems can be interpreted as consequences of unresolved symmetries [71]. Breaking these symmetries—through multi-view configurations, motion, or polarization diversity—is essential for accessing otherwise indistinguishable degrees of freedom. In the quotient-space framework, this corresponds to ensuring that the projection $\mathbf{P}_\perp$ is well-defined and that the trajectory excites its complement [72].

*Resource trade-offs in trajectory design.* The TDI framework makes explicit that different diversity mechanisms are not interchangeable in terms of resource cost [40, 41]. For example, increasing spatial aperture (e.g., longer baseline in bistatic radar, larger array in MIMO) increases TDI by improving angular resolution, but incurs physical footprint and hardware costs. Increasing frequency diversity (bandwidth) improves range resolution and geometric sensitivity, but consumes spectral resources and may increase hardware complexity. Increasing viewpoint variation through platform motion achieves similar angular diversity as spatial aperture, but trades off time and energy for physical movement [15, 17].

The TDI does not prescribe which mechanism is optimal—that depends on task-specific constraints—but it provides a common language for comparing how different diversity mechanisms contribute to excitation of the quotient space [59].

*Observability is not an intrinsic property of sensing hardware, but an emergent property of sensing trajectories interacting with physical symmetry and information geometry on the quotient space. In practice, performance gains arise not from increasing data volume, butfrom introducing new dimensions of variation—viewpoint, frequency, polarization—that expand the observable structure on the quotient space [53, 101, 15].*

*Summary: From Diagnostics to Design*

The design principles developed in this section can be summarized as follows:

- **The quotient space defines what is observable:** Symmetry reduction removes fundamental ambiguities, establishing $\Theta/\mathcal{G}$ as the intrinsic state space for sensing design [72];
- **The TDI diagnoses excitation:** The smallest eigenvalue of the projected Fisher information reflects the weakest excited observable direction. $\mathcal{D} = 0$ indicates structural blind spots; $\mathcal{D} \approx 0$ indicates ill-conditioning; $\mathcal{D}$ sufficiently large indicates well-excited, stable estimation [7];
- **Five observable signatures provide a complete health assessment:** Geometry–material decoupling, information growth, configuration diversity, cross-configuration agreement, and physics–data consistency each reveal different aspects of the underlying information geometry [43];
- **Failure modes map to signature violations:** Unresolved symmetry, imbalanced excitation, false identifiability, and physics–data mismatch each produce distinct diagnostic signatures that guide corrective action;
- **Trajectory design is information shaping on the quotient space:** Effective sensing requires coordinated diversity across multiple control dimensions to ensure all directions in $\Theta/\mathcal{G}$ are adequately excited [40, 59];
- **Observability is an engineered outcome:** The same hardware can yield different observability under different trajectories, and different architectures can achieve comparable observability when they induce similar information geometry on $\Theta/\mathcal{G}$ [41].

*Observability is not a fixed property of a sensing system, but an engineered outcome shaped by sensing actions and model fidelity. The five observable signatures provide the diagnostic language for this engineering process.*

## VII. Relation to Existing Paradigms

To clarify the contribution and avoid overclaiming, this section positions observability engineering within the broader landscape of sensing, inference, control, and learning. The proposed perspective is not intended to replace existing paradigms. Instead, it organizes them within a staged framework that links physical interaction, trajectory-induced information geometry, quotient-space identifiability, cross-system information equivalence, and observable world models.

As summarized in Fig. 1 and Table X, existing paradigms can be viewed as emphasizing different parts of this progression. Estimation theory studies inference accuracy under observation models; control-theoretic observability studies state reconstructability from measurements over time; experimental design studies measurement selection; information geometry characterizes local statistical distinguishability; and representation learning concerns the construction of task-relevant internal models [43, 37, 40]. Observability engineering connects these viewpoints by treating sensing as trajectory-driven information generation on an observable quotient space.

The framework developed throughout this paper is grounded in electromagnetic sensing governed by Maxwell's equations, with analysis primarily conducted under a high-frequency regime where geometrical optics provides a valid backbone approximation [44, 49, 45]. The results should therefore be interpreted within electromagnetic sensing contexts, rather than as a fully modality-agnostic abstraction. This scope is intentional: it preserves physical grounding while enabling general insights for practical sensing systems.

*1) Classical Radar Theory and Inverse Problems:* Classical radar theory establishes resolution limits based on bandwidth, aperture, waveform structure, and SNR under fixed sensing configurations [3, 4, 5]. The present framework inherits wave-based forward models derived from Maxwell's equations and the interpretation of resolution as a form of distinguishability under given sensing conditions. However, the proposed perspective further interprets resolution as a manifestation of trajectory-induced information geometry rather than as a fixed hardware property alone [43]. This reframes resolution limits from passive sensing constraints to quantities that can be actively shaped through sensing design [15, 17]. Hardware-level waveform optimization, clutter modeling, and implementation-specific radar engineering remain outside the present scope.

Inverse problems study the recovery of parameters from measurements under a fixed forward operator $H$ [119, 118, 64]. The present framework inherits the inverse-mapping viewpoint and the concept of ill-posedness. However, observability engineering emphasizes that the sensing process itself can shape the operator $H(u;\theta)$ through controllable sensing actions [40, 59]. In this perspective, sensing design modifies the inverse problem before inference is performed. Ill-posedness is therefore not only a property of the inverse solver, but also a consequence of sensing trajectory design [63]. Numerical solvers and regularization algorithms are not addressed here.

*2) SLAM, Robotics, and Experimental Design:* SLAM and robotics analyze which states are observable under motion and sensing constraints [31, 120, 113]. The present framework inherits the interpretation of observability as a determinant of estimability and the importance of motion diversity [28, 32]. However, observability engineering extends this perspective by formulating observability directly on quotient spaces, yielding a symmetry-aware characterization that applies beyond pose estimation and robotics-specific parameterizations [71, 72]. This suggests that observability is fundamentally a property of the sensing process rather than of a particular coordinate representation [37, 38]. Full SLAM pipelines, real-time implementations, and deployment issues are outside the present scope.

Optimal experimental design studies how measurements should be selected to maximize parameter information [40, 41, 121]. The present framework inherits Fisher information as a design criterion, but extends it by performing design directly on the observable quotient space and by enabling cross-system comparison through information correspondence [43, 41]. Different sensing architectures can therefore be compared through the information they generate rather than through hardware descriptors alone. Algorithmic optimality and finite-sample performance guarantees are not addressed here.

*3) Information Geometry and World Models:* Information geometry interprets Fisher information as a Riemannian metric on statistical models [43, 85, 122]. The present framework inherits the view of geometry as a measure of local statistical distinguishability. However, observability engineering further emphasizes that this geometry can be physically induced by controllable sensing trajectories and projected onto observable quotient spaces. This connects abstract information geometry to concrete sensing design decisions [105]. General manifold theory outside sensing applications is not considered.

Representation learning and world-model approaches seek to construct task-relevant quotient-space world models from data, often through deep learning architectures that compress high-dimensional observations into low-dimensional latent representations [111, 112, 123]. The present framework inherits the insight that quotient-space world models should be adaptive and interaction-driven rather than entirely hand-designed. However, the quotient-space formulation developed here provides a physics-grounded prior for representation learning. A well-structured world model should satisfy the invariance condition

$$\Phi(\theta) = \Phi(g \cdot \theta), \quad \forall g \in \mathcal{G},$$

whenever the observation distribution remains invariant under the symmetry group $\mathcal{G}$. This directly connects geometric deep learning [111] to the physical symmetries of sensing systems.

Moreover, the projected Fisher information $\mathbf{I}_Q$ defines a local Riemannian metric that can guide representation learning: directions with high Fisher information require higher representational capacity, while directions with near-zero information can be compressed or ignored [43, 105]. Representation learning and world modeling are therefore not separate from observability engineering; they become data-driven realizations of the quotient-space structure and information geometry developed throughout Stages 2–4.

Implementation of learnable world models and empirical validation on real sensing systems remain outside the present scope. The contribution here is to clarify what structural properties a representation should satisfy, rather than how such representations should be learned from data. End-to-end learning of observable representations and large-scale empirical evaluation remain important open directions.

Together, these connections suggest that observability engineering is not a competing paradigm, but a unifying lens through which existing methods can be understood, compared, and extended. More fundamentally, each field captures a different projection of the same underlying information-generation process. The contribution of the present framework is therefore not to replace these paradigms, but to show how they fit together into a coherent engineering structure.

TABLE X
POSITIONING EXISTING PARADIGMS WITHIN THE OBSERVABILITY-ENGINEERING FRAMEWORK. THE PURPOSE IS NOT TO REPLACE THESE PARADIGMS, BUT TO SHOW WHICH STAGES OF THE PROPOSED FRAMEWORK THEY PRIMARILY EMPHASIZE.

| Paradigm | Primary Focus | Primary Connection to This Framework |
|---|---|---|
| Radar and array sensing | Resolution, aperture, waveform, and SNR limits | Primarily Stage 1: physics-constrained generation of information geometry through Maxwell observation families, sensing trajectories, and Fisher information (Section II). |
| Inverse problems | Recovering parameters from a forward observation model | Connects Stage 1 and Stage 2: the forward operator defines sensitivity, while quotient-space identifiability clarifies which parameter directions are intrinsically observable (Sections II–III). |
| Control-theoretic observability | State reconstructability from measurements over time | Related to Stage 2: symmetry-reduced observability separates intrinsic non-identifiability from trajectory-induced insufficient excitation (Section III). |
| SLAM and active perception | State estimation, mapping, and action selection under uncertainty | Connects Stage 2 and the operational diagnostics: sensing actions should excite quotient-identifiable directions and update observable representations (Sections III and VI). |
| Optimal experimental design | Selecting measurements to improve estimation accuracy | Related to Stage 1 and the operational diagnostics: measurement selection is interpreted as trajectory design for shaping Fisher information on observable directions (Sections II and VI). |
| Information geometry | Local statistical distinguishability and metric structure | Provides the mathematical language for Stage 1–3: Fisher information characterizes local distinguishability induced by sensing trajectories (Section II). |
| Representation learning and world models | Constructing task-relevant quotient-space world models | Related to Stage 4: world models should encode observable structure on $\Theta/G$, rather than redundant ambient parameterizations (Section V). |

For practitioners, this means that insights from radar, SLAM, inverse problems, experimental design, and representation learning can be combined within a single framework centered on quotient-space observability and trajectory-induced information geometry.

In the present work, this pipeline is instantiated explicitly within electromagnetic sensing, where Maxwell-consistent observation models and their high-frequency structure determine the resulting information geometry [44, 45, 49]. The same principles, however, may extend to other sensing modalities whenever analogous physical and symmetry structures exist.

## VIII. Conclusion and Outlook

This article has developed a unified structural interpretation of observability for modern sensing systems. The central message is that observability is not a property of measurement alone. It emerges jointly from physical interaction, information generation, symmetry-constrained identifiability, and quotient-space world models.

Within this viewpoint, sensing actions shape the information geometry that governs distinguishability [59, 43]; symmetry organizes latent physical parameters into observable equivalence classes [71, 72]; heterogeneous sensing architectures may be related through quotient-level correspondence [41, 40]; and adaptive sensing requires quotient-space world models whose state structure remains consistent with the observable geometry induced by the sensing process [31, 32].

This perspective does not define a new isolated sensing paradigm. Rather, it provides a structural synthesis of principles that have progressively appeared across radar, estimation theory, active perception, robotics, integrated sensing and communications, and embodied autonomous systems [3, 7, 28, 31, 24, 124]. In doing so, it offers a common engineering language for reasoning about what can be observed, what cannot be distinguished, how heterogeneous sensing architectures may be compared, and how representational or adaptive failure modes may be diagnosed [94, 53, 101].

More fundamentally, classical radar theory, inverse problems, SLAM, experimental design, and information geometry can all be interpreted as different manifestations of a common principle: information is generated by sensing actions and structured by observability on the quotient space. This unifying view connects resolution, identifiability, and representation within a single design-oriented framework, where sensing, inference, and decision-making are no longer separate stages, but tightly coupled components of an observability-driven loop [28, 32].

Within this framework, world modeling emerges naturally as the construction of representations on the observable quotient space $\Theta/\mathcal{G}$, where symmetry-induced ambiguities are removed and uncertainty is organized through the induced information geometry. Consistency across sensing configurations is therefore understood not as agreement in parameterization, but as convergence toward the same observable structure [31, 110, 115].

As sensing systems become increasingly autonomous, distributed, and multi-modal, the ability to reason jointly about information generation, observable structure, and representation-level consistency will become a foundational requirement rather than a specialized abstraction. This structural viewpoint therefore provides a principled basis for the analysis, design, and operational diagnosis of future adaptive sensing systems.

Several research directions emerge naturally from this perspective.

*1) Information-Driven Trajectory Design:* The geometry induced by sensing actions suggests a control problem: selecting trajectories that maximize information about observable states. This connects observability engineering to active perception, optimal experimental design, and reinforcement learning, particularly in non-Gaussian and partially observable settings where the quotient structure may evolve [28, 32, 125, 126]. Extending the framework beyond local Fisher geometry toward global information measures, such as Kullback–Leibler divergence, Wasserstein distance, and Bayesian decision risk, remains an important direction for future work [73, 74, 42].

*2) Multi-Agent and Distributed Sensing:* Information correspondence provides a basis for coordinating multiple sensing agents. Rather than duplicating measurements, agents can distribute effort across complementary regions of the information space, enabling efficient and scalable information acquisition under heterogeneous sensing capabilities [33, 58, 32].

*3) Integrated Sensing and Communications (ISAC):* The framework offers a lens for jointly analyzing sensing and communication, where Fisher information governs estimation and mutual information governs data transmission [106]. Understanding their interaction may lead to unified design principles for systems that simultaneously sense and communicate, with recent information-theoretic analyses providing initial characterizations of fundamental tradeoffs [24, 23, 127, 128].

*4) Hybrid Physics-Informed Learning:* The decomposition of sensing models into physics-based backbones and data-driven residuals suggests hybrid architectures that combine robustness and flexibility [129, 130]. A key direction is ensuring that learned representations respect symmetry and identifiability structures implied by observability, an area where geometric deep learning offers promising tools [111, 112, 123, 131].

*5) Equivariant and Geometric Deep Learning:* The quotient-space formulation naturally connects to geometric deep learning, where group-equivariant representations enforce invariance to physical symmetries [111, 112, 123]. This connection suggests that observability-aware representations can be learned efficiently using equivariant neural network architectures, opening new avenues for integrating physical priors into data-driven sensing systems.

Taken together, these directions point toward a unified paradigm in which *physics defines what can be observed, information geometry quantifies how well it can be observed, and system design determines how information is generated and used*. Within this paradigm, sensing, learning, and decision-making become tightly coupled components of a single closed-loop process.

Ultimately, the significance of observability engineering lies not in introducing a new technique, but in revealing a common structure that governs how sensing systems generate, organize, and exploit information.

In this sense, sensing is not merely about observing the world, but about shaping the space of what can be observed.

## Acknowledgment

The authors would like to thank the Editor-in-Chief and the editorial team of *Proceedings of the IEEE* for the opportunity to contribute this invited tutorial article. The authors also acknowledge the use of an AI-based language assistant for editorial suggestions and language polishing during the preparation of this manuscript. All scientific concepts, theoretical developments, and technical conclusions presented in this work were developed and verified by the authors.

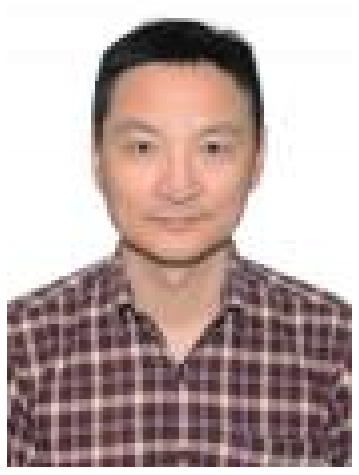

**Pin-Han Ho** started working in the Department of Electrical and Computer Engineering, University of Waterloo, Canada, as a faculty member since 2002. He is currently a full Professor in the same department and also with Shenzhen Institute for Advanced Study, China as a senior researcher. His current research interests cover a wide range of topics on information fusion, multi-agent sensing, and AI-empowered system engineering. He is ranked as Fellow IEEE and Fellow AAIA, and a licensed PEO.

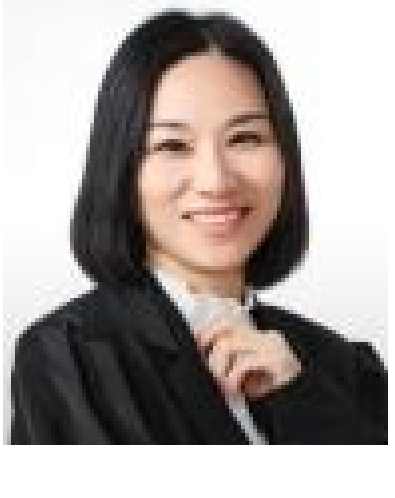

**Limei Peng** is with the School of Computer Science and Engineering, Kyungpook National University (KNU), Daegu, South Korea, and Shenzhen Institute for Advanced Study, University of Electronic Science and Technology of China. Her research interests include cloud computing, fog computing, data center networks, Internet of Things (IoT)/Internet of Vehicles (IoV), and 5G/6G communications networks.